\documentclass[12pt,preprint]{aastex}

\usepackage{amssymb}
\usepackage{natbib}
\usepackage{epsfig}

\newcommand{\Msol}{\ensuremath{\mathrm{M_{\odot}}}}

\def\R200{\ensuremath{R_{\mathrm{200}}}}

\newcommand{\Zsol}{\ensuremath{\mathrm{Z_{\odot}}}}

\newcommand{\egc}{{\it e.g.}}  
\newcommand{\etal}{{\it et al.\thinspace}}

\newcommand{\ie}{{\it i.e.\ }}

\newcommand{\ASCA}{\emph{ASCA}\ }
\newcommand{\Chandra}{\emph {Chandra}\ }

\newcommand{\ROSAT}{\emph{ROSAT}\ }
\newcommand{\XMM}{\emph{XMM-Newton}\ }

\newcommand{\BeppoSAX}{\emph{BeppoSAX}\ }
\newcommand{\MEKAL}{\textsc{MeKaL}\ }

\newcommand{\chisq}{\ensuremath{\chi^2}}

\newcommand{\nm}{\mbox{\ensuremath{\mathrm{~\nm}}}}
\newcommand{\cm}{\mbox{\ensuremath{\mathrm{~cm}}}}
\newcommand{\m}{\mbox{\ensuremath{\mathrm{~m}}}}
\newcommand{\km}{\mbox{\ensuremath{\mathrm{~km}}}}

\newcommand{\kpc}{\mbox{\ensuremath{\mathrm{~kpc}}}}
\newcommand{\Mpc}{\mbox{\ensuremath{\mathrm{~Mpc}}}}
\newcommand{\s}{\mbox{\ensuremath{\mathrm{~s}}}}
\newcommand{\ks}{\mbox{\ensuremath{\mathrm{~ks}}}}

\newcommand{\keV}{\mbox{\ensuremath{\mathrm{~keV}}}}
\newcommand{\erg}{\mbox{\ensuremath{\mathrm{~erg}}}}

\newcommand{\degree}{\ensuremath{\mathrm{^\circ}}}
\newcommand{\arcm}{\ensuremath{\mathrm{^\prime}}}
\newcommand{\arcs}{\arcm\hskip -0.1em\arcm}
\newcommand{\dotarcs}{\,\rlap{\hbox{$\mathrm{^\prime\hskip -0.1em^\prime}$}}{\hbox{$.$}}\,}
\newcommand{\dotsec}{\,\rlap{\hbox{$\mathrm{^s}$}}{\hbox{$.$}}\,}

\newcommand{\cmsq}{\ensuremath{\mathrm{\cm^2}}}

\newcommand{\pcc}{\ensuremath{\mathrm{\cm^{-3}}}}
\newcommand{\pcmsq}{\mbox{\ensuremath{\mathrm{~cm^{-2}}}}}
\newcommand{\pMpc}{\ensuremath{\mathrm{\Mpc^{-1}}}}
\newcommand{\ps}{\ensuremath{\mathrm{\s^{-1}}}}

\newcommand{\ergps}{\ensuremath{\mathrm{\erg \ps}}}
\newcommand{\flux}{\ensuremath{\mathrm{\erg \ps \pcmsq}}}
\newcommand{\kmps}{\ensuremath{\mathrm{\km \ps}}}
\newcommand{\kmpspMpc}{\ensuremath{\mathrm{\km \ps \pMpc\,}}}

\newcommand{\LCDM}{$\Lambda$CDM~}

\shorttitle{X-ray analysis of clusters ClJ1113.1$-$2615 and ClJ0152.7$-$1357}
\shortauthors{[B.J. Maughan \etal}

\received{2002 October 1}
\begin{document}

\title{Chandra X-ray analysis of the massive high-redshift galaxy clusters ClJ1113.1$-$2615 and ClJ0152.7$-$1357}

\author{B. J. Maughan}
\affil{School of Physics and Astronomy, University of Birmingham, Edgbaston, Birmingham, B15 2TT, UK}
\email{bjm@star.sr.bham.ac.uk}
\and
\author{L. R. Jones}
\affil{School of Physics and Astronomy, University of Birmingham, Edgbaston, Birmingham, B15 2TT, UK}
\and
\author{H. Ebeling}
\affil{Institute for Astronomy, 2680 Woodlawn Drive, Honolulu, HI 96822, USA}
\and
\author{E. Perlman\altaffilmark{1}}
\altaffiltext{1}{Department of Physics \& Astronomy, Johns Hopkins University, 3400 North Charles Street, Baltimore, MD 21218, USA}
\affil{Joint Centre for Astrophysics, Physics Department, University of Maryland,  Baltimore County, 1000 Hilltop Circle, Baltimore, MD 21250, USA}
\and
\author{P. Rosati}
\affil{European Southern Observatory (ESO), Headquarters, Karl-Schwarzschild-Strasse 2, Garching D-85748, Germany}
\and
\author{C. Frye}
\affil{Joint Centre for Astrophysics, Physics Department, University of Maryland,  Baltimore County, 1000 Hilltop Circle, Baltimore, MD 21250, USA}
\and
\author{C. R. Mullis}
\affil{European Southern Observatory (ESO), Headquarters, Karl-Schwarzschild-Strasse 2, Garching D-85748, Germany}

\newpage
\begin{abstract}
{\small We present an analysis of \Chandra observations of two high-redshift
clusters of galaxies, ClJ1113.1$-$2615 at $z=0.725$ and
ClJ0152.7$-$1357 at $z=0.833$. We find ClJ1113.1$-$2615 to be
morphologically relaxed with a temperature of
$kT=4.3^{+0.5}_{-0.4}\keV$ and a mass (within the virial radius) of
$4.3^{+0.8}_{-0.7}\times10^{14}M_\odot$. ClJ0152.7$-$1357, by
contrast, is resolved into a northern and southern subcluster, each
massive and X-ray luminous, in the process of merging. The
temperatures of the subclusters are found to be
$5.5^{+0.9}_{-0.8}\keV$ and $5.2^{+1.1}_{-0.9}\keV$ respectively, and
we estimate their respective masses to be
$6.1^{+1.7}_{-1.5}\times10^{14}M_\odot$ and
$5.2^{+1.8}_{-1.4}\times10^{14}M_\odot$ within the virial radii. A dynamical analysis of the
system shows that the subclusters are likely to be gravitationally
bound. If the subclusters merge they 
will form a system with a mass similar to that of the Coma cluster. Two-dimensional modelling of the
X-ray surface brightness reveals excess emission between the
subclusters; suggestive, but not conclusive evidence of a shock front.

We make a first attempt at measuring the cluster M-T relation at z$\approx$0.8,
and find no evolution in its normalisation, supporting the
previous assumption of an unevolving M-T relation when constraining
cosmological parameters from cluster evolution studies.
A comparison of the cluster properties with those of nearby
systems also finds  little or no evolution in the L-T relation, 
the gas fraction-T relation,  the $\beta$-T relation or the metallicity.
These results suggest that, in at least some massive
clusters, the hot gas was in place, and containing its metals, at $z\approx0.8$, and thus that
they were assembled at redshifts significantly higher than $z=0.8$, as
predicted in low $\Omega_M$ models. 

We also highlight the need to correct for the degradation of the $Chandra$ 
ACIS low energy quantum efficiency in high-redshift cluster studies when 
the low energy absorption is often assumed to be the Galactic value, rather than 
measured.}
\end{abstract}


\section{Introduction} \label{sect:intro}
Detailed X-ray studies of distant galaxy clusters have only recently
become possible thanks to powerful new instruments aboard the \Chandra
and \XMM satellites. These clusters represent the largest bound systems
in the universe, and the study of their properties, both individually
and as a population, allows different cosmological and structure
formation models to be tested and constrained.

Many cosmological models describe the formation of structure in the
Universe in terms of a hierarchical model, in which initial density
perturbations (the distribution of which is usually assumed to be
Gaussian) in the early universe grow under gravity, eventually forming
the massive structures seen today. In this scenario, less massive
systems form first. Then, through a combination of merging and the
ongoing collapse of the slightly less dense surrounding regions,
larger systems of galaxies, groups, and clusters of galaxies are
formed. The immediately observable, luminous components of these
systems act as tracers of the underlying distribution of dark matter,
which comprises the majority of the mass in the universe. While the
basic elements of this scenario are widely accepted, many important
details, specifically the epoch and mode of cluster assembly, remain
largely unknown. Massive clusters are especially useful for
quantitative studies because their abundance at high redshifts is very
sensitive to the matter density of the Universe, their evolution is
predicted to be strongest, and -- being the most luminous systems --
they are the ones that are observationally most accessible.

The X-ray properties of galaxy clusters have been exploited as
cosmological probes for many years, using, for example, their
temperatures \citep[\egc][]{hen91,hen00}, luminosities
\citep[\egc][]{jon98a,bor01}, and more recently, gas mass fractions
\citep{all02a,ett03}. Since the launch of \Chandra and {\em XMM-Newton}, these
properties of clusters can be much more accurately measured out to
higher redshifts, enabling us to constrain cosmological parameters to
much higher precision. \Chandra observations in particular have also
found new and sometimes unexpected features in clusters, such as cold
fronts \citep[\egc][]{mar00c}, radio cavities \citep[\egc][]{mcn01}, and shock fronts \citep[\egc][]{mar02}. The detailed study of
individual clusters is an important step to compiling samples of
high-redshift clusters with well constrained properties.

Both clusters discussed here, ClJ1113.1$-$2615 and ClJ0152.7$-$1357,
were discovered in the Wide Angle \ROSAT Pointed Survey (WARPS:
\citet{sch97}); ClJ0152.7$-$1357 was also discovered independently in
the RDCS \citep{ros98} and SHARC \citep{rom00} surveys. However, due
to their high redshifts and the limitations of previous instruments,
relatively little is known about their properties. Optical
spectroscopy was performed on galaxies in the region of the \ROSAT
X-ray sources to measure their redshift and to confirm their cluster
status \citep{ebe00,per02}. The \ROSAT observations of
ClJ0152.7$-$1357 have been analysed in some detail \citep{ebe00}, and
the cluster has also been the subject of both \BeppoSAX observations
\citep{del00} and Sunyaev-Zel'dovich effect imaging \citep{joy01}. The
recent \Chandra observations of these clusters, presented here, both
complement and extend this previous work, and the results of the
different observations are compared later. These observations, along
with \Chandra and \XMM observations of other high-redshift clusters,
will form the basis of a study of the evolution of the X-ray
temperature function, thereby constraining the value of $\Omega_{M}$.

Section \ref{sect:dataprep} describes the general data preparation and
analysis methods applied to the \Chandra data for both
objects. Sections \ref{sect:J1113} and \ref{sect:J0152} detail the
results obtained for ClJ1113.1$-$2615 and ClJ0152.7$-$1357
respectively, before we summarise and discuss these results in section
\ref{sect:discussion}.

We perform our analysis assuming two different cosmological models: a
\LCDM model with $\Omega_{M}=0.3$, $\Omega_\Lambda=0.7$ and
$H_0=70\kmpspMpc$, and an Einstein-de Sitter model with
$H_0=50\kmpspMpc$ and $\Omega_M=1$ $(\Omega_\Lambda=0)$. The Einstein-de Sitter model was chosen to allow comparisons with earlier work. Unless
otherwise stated, results are quoted for the \LCDM cosmology,
and all errors are quoted at the $68\%$ confidence level.

\section{Data Preparation and Analysis} \label{sect:dataprep}
Our two targets were observed with the ACIS-I array of the \Chandra
X-ray Observatory, and the same standard data preparation steps were
followed for both clusters as described below. ClJ0152.7$-$1357 was observed on  $2000$ August $8$ (ObsID 913) and ClJ1113.1$-$2615 was observed on $2000$ August $13$ (ObsID 915).

Both data sets were reprocessed by the standard pipeline software, removing bad
events, and excluding \ASCA grades 1, 5, and 7 (corresponding to
diagonal split, ``L''-shaped split, and 3-pixel horizontal split
events). The data were then corrected to the appropriate gain map, and
for any known aspect offsets. A background lightcurve with time bins of $400\s$ was produced and analysed. The
lightcurve of ClJ0152.7$-$1357 showed only brief periods of enhanced
background, while there was a significant flaring event in the
observation of ClJ1113.1$-$2615. All time intervals during which the
background was $>3\sigma$ from the mean level were removed, leaving a
useful time of $34\ks$ for ClJ0152.7$-$1357 and $88\ks$ for
ClJ1113.1$-$2615. Point sources were identified with the CIAO wavelet detection algorithm, wavdetect.

\subsection{Spectral Analysis}
Spectra were extracted from circular regions centred on the clusters'
X-ray centroids, within a radius chosen from a preliminary radial profile of
the emission to yield good signal-to-noise ratios (snr). These radii
were typically $\sim50\arcs$ and are hereafter referred to as the
spectral radii ($r_s$) of the clusters. The Redistribution Matrix File
(RMF), which relates the incident photon energy to the output channel
energy of the instrument, and the Ancillary Response File (ARF), which
describes the dependence of effective area upon energy, both vary
spatially across the detector. Since our extended sources cover
regions described by several different RMFs and ARFs, a mean of each
of these files, weighted by the source counts, was produced and used
in the spectral analysis.

Background spectra were extracted from annuli surrounding the
clusters, on the same CCD, excluding any point sources and pixels near
the chip edges. An alternative method of obtaining background spectra
from blank-sky datasets \citep{mar00b} was also examined. This method
has the advantage that the background spectra are extracted from the
same region of the CCD as the source, thus eliminating potential
systematic errors caused by the spatial variations of both the energy
response and the effective area across the chip. Changes in the
quiescent background between the time of the blank-sky observation and
that of the science observation being analysed can be, approximately,
corrected for by normalising the blank-sky data appropriately. The
method remains, however, vulnerable to temporal variations in the
spectrum of the particle background, and also cannot easily account
for the known strong directional variation of the Galactic soft X-ray
emission. A comparison of both methods showed small $(\sim1\sigma)$
differences in the values of the best-fit parameters. Given the
compactness of our targets on an angular scale, we felt that the
former method of using a background spectrum created from the same
observation as the target involved approximations that were better
understood, and so this method was used in all of the analyses
detailed below.

It has recently become apparent that there has been a continuous degradation in the ACIS low energy quantum efficiency (QE) since the launch of \emph{Chandra}. This is believed to be due to a build up of hydrocarbons on the optical filter, or the CCDs, which has the effect of introducing extra absorption below around $1\keV$. In the context of our data, if unmodelled, this absorption leads to an overestimate in cluster temperatures. To account for this, an XSPEC absorption model ACISABS has been developed\footnote{http://www.astro.psu.edu/users/chartas/xcontdir/xcont.html}, and the spectra presented here were reanalysed with this model. The effect of the QE correction is examined in Appendix \ref{app1}.

The resulting spectra were fit with an absorbed \MEKAL plasma emission
model \citep{kaa93,lie95} in both Sherpa and XSPEC \citep{arn96}, with completely
consistent results. The normalisation of the \MEKAL component was used
to calculate the central density of the intra-cluster gas, $\rho_0$,
as follows. The \MEKAL normalisation in XSPEC is defined as
\begin{eqnarray}
\label{eqn:Nmek}
N_{mek} & = & \frac{10^{-14}}{4\pi\,D_a^2\,(1+z)^2}\int n_e\,n_H\,dV \mathrm{cm^{-5}},
\end{eqnarray}
where $D_a$ is the angular diameter distance in cm, and $n_e$ and
$n_H$ are the central number densities of electrons and hydrogen
respectively. Under the assumption that the
cluster gas has a spherically symmetric density profile described by a
$\beta$-profile \citep{cav76} (the parameters $r_c$ and $\beta$ of
which are derived from the data as described in the following
section), we have
\begin{eqnarray}
\label{eqn:bprof}
\rho(r) & = & \rho_0\left(1+\left(\frac{r}{r_c}\right)^2\right)^{-3\beta/2}.
\end{eqnarray}
Considering that the spectrum is taken from a circular region of
radius $r_s$, viewing the sphere in projection, the integral of
Eqn. \ref{eqn:Nmek} can be expressed as
\begin{eqnarray}
\int1.17\,\rho^2\,dV & = & 1.17\left(\int^{\infty}_{0}\rho(r)^2 \,4\pi \,r^2dr - \int^{\infty}_{r_s}\rho(r)^2 \,4\pi \,r^2\left(1-\left(\frac{r_s}{r}\right)^2\right)^{0.5}dr\right).
\end{eqnarray}
Here we assume $n_e=1.17n_H$. The central gas density $\rho_0$ can then be derived from the \MEKAL normalisation via Eqn. \ref{eqn:Nmek}.

\subsection{Spatial Analysis}\label{sect:spatial}
The spatial analysis of the cluster X-ray emission was performed in
the $0.5-5\keV$ energy range within which the spectra, extracted as detailed
in the preceding section, showed to feature the highest snr. The
spectra were also used to produce spectrally weighted exposure
maps\footnote{For \Chandra data, the standard analysis software produces an exposure map that is an image of the
effective exposure time in units of $\s\cmsq$ in the detector plane,
taking into account the vignetting of the telescope, CCD gaps, the
effective area of the telescope, and the dithering of the
satellite. By producing a spectrally weighted exposure map, in effect
combining exposure maps in several energy ranges weighted by the
number of counts in each range, one also accounts for the energy
dependence of the vignetting and of the effective area.} for the
imaging analysis.  

Two kinds of models of the cluster emission were employed -- a
one-dimensional (1D) model of the radial surface-brightness profile
assuming spherical symmetry, and a two-dimensional (2D) model of the
emission which is fit to an image (see \citet[\egc][]{bir91} for a discussion of 2D fitting). When producing a radial profile it
is desirable to have $\ga20$ counts in each radial bin, so that
Gaussian errors may be assumed\footnote{For significances below
$3\sigma$, $n\ge 20$ ensures that the Gaussian approximation is
accurate to better than 30 (10) \% for the lower and upper limits,
respectively, when compared to the exact Poisson errors
\citep{ebe02}.}. When dividing a raw-counts image by an exposure map,
the latter was normalised to its value at the cluster centroid, so
that the exposure-corrected image maintains, as much as possible, the
photon statistics of the recorded counts image. Since our exposure
maps vary by typically no more than $\approx5\%$ across the region of
interest, this approach should not noticeably affect the accuracy of
our results or their errors. 

A second concern in any imaging analysis is the impact of the point
spread function (PSF) of the telescope. In our 1D analysis we do not
account for the PSF at all, because the \Chandra PSF FWHM is smaller
than the typical bin size in the radial profiles, so should have
little effect.  The 2D method deals more elegantly with both the
exposure map and the PSF: the 2D model is convolved with a PSF model
generated for the energy and off-axis position of the cluster, and
then multiplied by the exposure map, before being compared with the
original, unflattened image. This has the advantage of preserving the
Poisson statistics of the observed image. The disadvantage of the 2D
approach is that, since each image pixel contains very few (or zero)
counts, we must use the C statistic \citep{cas79} to find the best
fit. Unlike the $\chisq$ statistic, the value of the C statistic gives
no measure of the absolute goodness of fit. The $\chisq$ statistic may
be used though when fitting a radial profile, provided the radial bins
contain sufficient counts. Thus the approach followed in our imaging
analysis was to derive the best fit parameters and their errors with
the 2D method, and use 1D profiles to check consistency and judge the
goodness-of-fit.

In general, we model the surface-brightness distribution of each
cluster, or subcluster, with a $\beta$-model given by
\begin{eqnarray}
S(r) & = & S_0\left(1+\left(\frac{r}{r_c}\right)^2\right)^{-3\beta+1/2},
\end{eqnarray}
where $S(r)$ is the surface brightness at radius $r$, $S_0$ is the
central surface brightness, $r_c$ is the core radius, and $\beta$
describes the slope of the profile at large radii. In the case of the
2D models, the eccentricity ($e$) and the rotation angle ($\theta$) of
the semi-major axis of the model relate the Cartesian (x,y)
coordinates of a point in the image to the radial coordinate.

The integrated properties of the cluster gas (\ie flux, luminosity and
mass) were then estimated at the virial radius by extrapolating their
values, measured within the spectral radius, using the measured model
profile. The virial radius of a cluster may be estimated using
self-similar scaling arguments, normalised to numerical
simulations. Here, we adopted the formalism of \citet{arn02a}, which
may be applied to a flat ($\Omega_0+\Lambda=1$) or open
($\Omega_0<1,\Lambda=0$) Universe:
\begin{eqnarray}
r_v & = & 3.80\,\beta_T^{1/2}\,\Delta_z^{-1/2}\,(1+z)^{-3/2} \nonumber \\
    &   & \times\left(\frac{\rm{k}T}{10\keV}\right)^{1/2}h_{50}^{-1}\Mpc. \label{virrad}
\end{eqnarray}
Where $\Delta_z = (\Delta_c\,\Omega_0)/(18\pi^2\,\Omega_z)$, $\beta_T
= 1.05$ is a normalisation, taken from \citet{evr96}, and the density
contrast $\Delta_c$ is given by
\begin{eqnarray}
\Delta_c & = & 18\pi^2 + 60\,(\Omega_z-1) - 32\,(\Omega_z-1)^2 
\end{eqnarray}
for an open Universe, and
\begin{eqnarray}
\Delta_c & = & 18\pi^2 + 82\,(\Omega_z-1) - 39\,(\Omega_z-1)^2 
\end{eqnarray}
for a flat Universe. The matter density at redshift $z$, $\Omega_z$,
is given by
\begin{eqnarray}
\Omega_z & = & \frac{\Omega_0\,(1+z)^3}{\Omega_0\,(1+z)^3+(1-\Omega_0-\Lambda)\,(1+z)^2+\Lambda}.
\end{eqnarray}

The total gravitating mass of the cluster can then be estimated if we
assume that the cluster is spherically symmetric, in hydrostatic
equilibrium, and is well described by an isothermal
$\beta$-profile. Adopting further a value of $0.59m_p$ for the mean
molecular weight of the gas, where $m_{\rm p}$ is the proton mass, we
can express the total mass of the cluster within a radius $r$ as
\citep{evr96},
\begin{eqnarray}
\label{egn:hydroeqm}
M(r) & = & 1.13\times10^{14}\beta\frac{T}{\keV}\frac{r}{\Mpc}\frac{(r/r_c)^2}{1+(r/r_c)^2}M_{\odot}.
\end{eqnarray}

\subsection{Treatment of Errors}
The errors of the measured fluxes (within the extraction regions) were
calculated from the fractional errors on the normalisation of the
spectrum; errors of all derived quantities were computed
self-consistently with a `quasi Monte-Carlo method' as follows. For
each measured parameter $T, \beta$, and flux, an asymmetric Gaussian
was computed, centred on the best-fit value and with the positive and
negative $\sigma$ widths given by the positive and negative $1\sigma$
errors found from the best fits. 100,000 values were selected at
random from these distributions to compute the distributions of each
dependent quantity, such as $L_X$ or $M$. The errors quoted for each
quantity are the range within which $\pm 34\%$ of the values
occurred. The core radius $r_c$ was fixed during this process. The errors on $\beta$ from the surface brightness fitting were computed with $r_c$ free to vary, to account for the correlation between these variables. The contributions of the errors on $r_c$ were found to be insignificant compared to those on $beta$, and so were ignored. This method highlights
the uncertainties involved in extrapolating the emission beyond the
spectral radius: the extrapolation is strongly dependent on the slope
of the surface-brightness profile. Lower values of $\beta$ correspond
to a shallower profile, with a smaller fraction of the emission being
directly measured, leading to larger uncertainties on the extrapolated
values, such as the scaled fluxes. The spread of values of $\beta$,
therefore, contributes significantly to the errors on the scaled
parameters.

\section{Cluster ClJ1113.1$-$2615} \label{sect:J1113}
\begin{figure*}
\begin{center}
\plotone{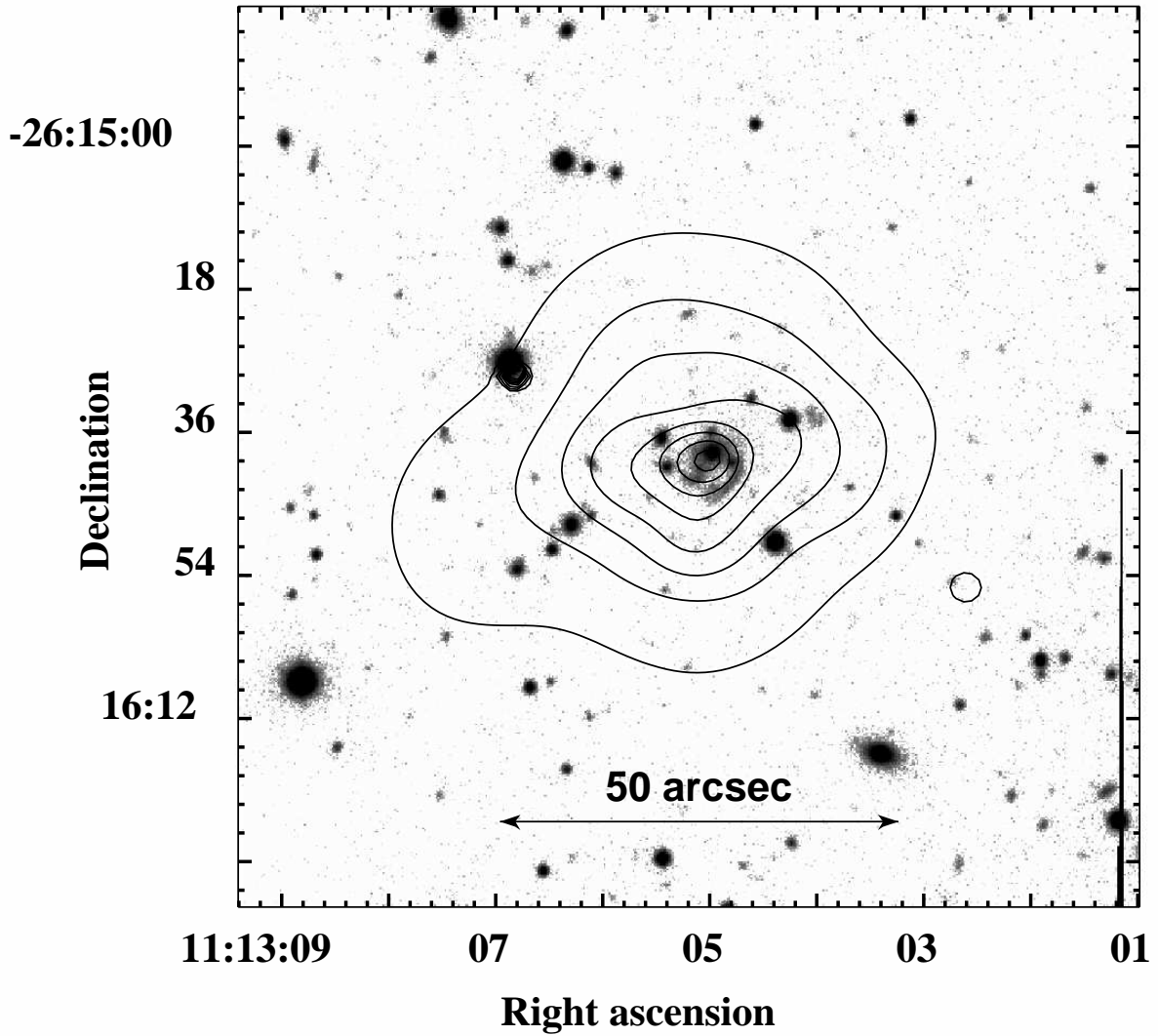}
\caption{\label{fig:j1113_overlay}Adaptively smoothed \Chandra X-ray
contours overlaid on a Keck II (LRIS) I-band image of ClJ1113.1$-$2615. The
contours were taken from an exposure corrected image, which was adaptively smoothed so that all features are significant at the $99\%$ level. The contours start at The contours start at
$0.2\mathrm{~counts ~pixel^{-1}}$, and are logarithmically spaced by a factor of $1.2$.}
\end{center}
\end{figure*}

This cluster was identified as a low-surface-brightness, extended
X-ray source in the WARPS survey \citep{per02}. It has a high redshift
of $z=0.725$ giving a scale of $7.2\kpc$ per arcsecond (\LCDM
cosmology). In Fig. \ref{fig:j1113_overlay}, contours of the X-ray
emission detected by \Chandra are overlaid on an optical I-band image,
obtained at the Keck II $10\m$ telescope. The inner X-ray contours appear smooth and fairly regular, indicating that, at least in the central regions, the
intra-cluster gas is in a relaxed state. This supports the assumption
of hydrostatic equilibrium which underpins the derivation of several
of the cluster's properties, such as the mass and virial radius. The
region shown in Fig. \ref{fig:j1113_overlay} contains one faint X-ray point
source approximately $25\arcs$  north-east of the cluster.

\clearpage

\clearpage

\begin{figure}
\begin{center}
\plotone{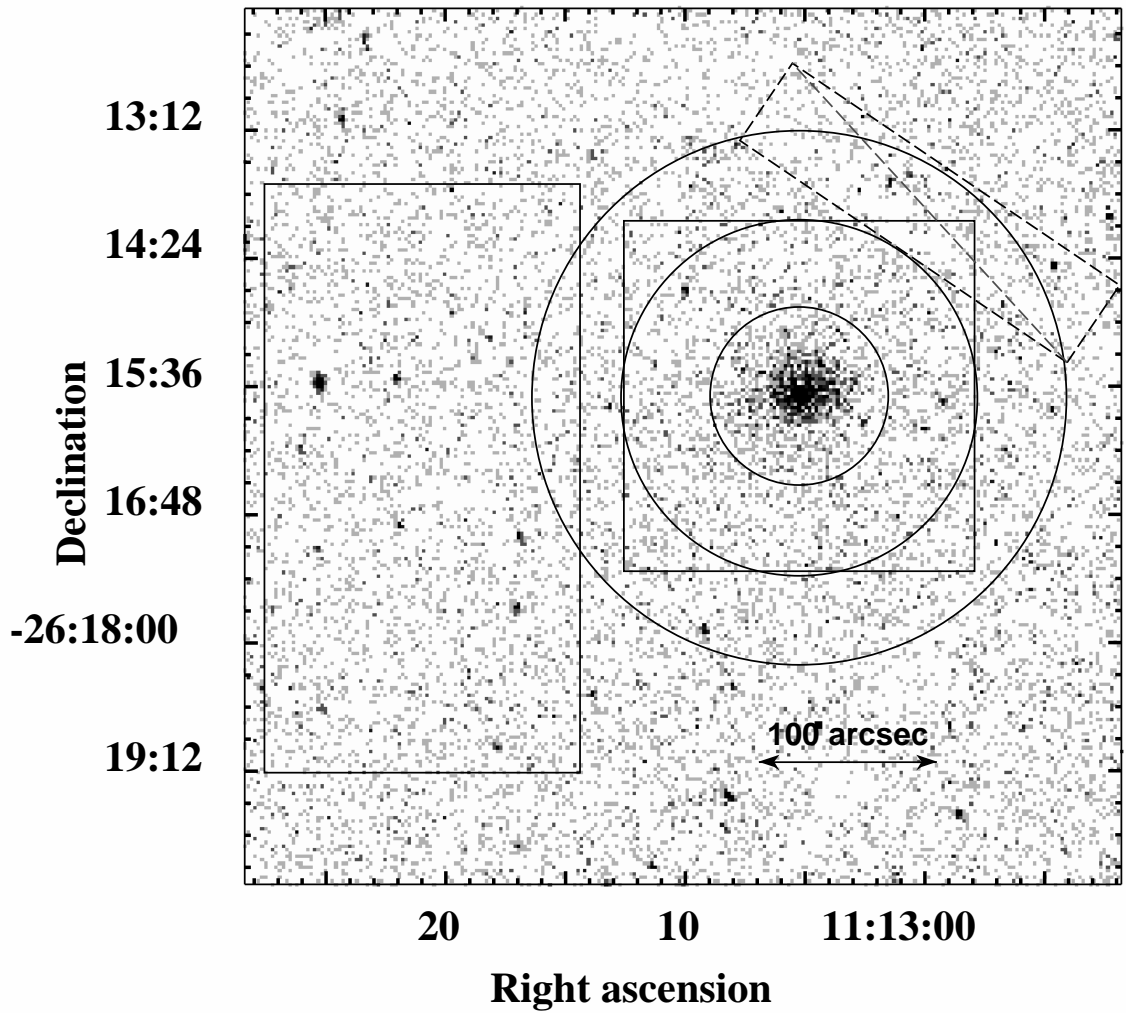}
\caption{\label{fig:j1113regions}X-ray image of ClJ1113.1$-$2615 with pixels of $1.968\arcs$. 
The regions used for imaging (rectangular) and spectral (circle and annulus) analysis are shown.
The dashed rectangular region was ignored because it was close to a CCD edge.}
\end{center}
\end{figure}

Fig. \ref{fig:j1113regions} indicates the regions used for the
extraction of products. The rectangular regions are the source and
background regions used in the imaging analysis, while the circle and
annulus are the source and background regions, respectively, used in
the spectral analysis. Note that the barred, dashed rectangular region
indicates data along the CCD edge that were excluded from the spectral
background. All point sources were excluded from the following
spectral and imaging analysis.

\clearpage

\clearpage

\subsection{Spectral results}\label{sect:j1113spectral}
\begin{figure}
\begin{center}
\plotone{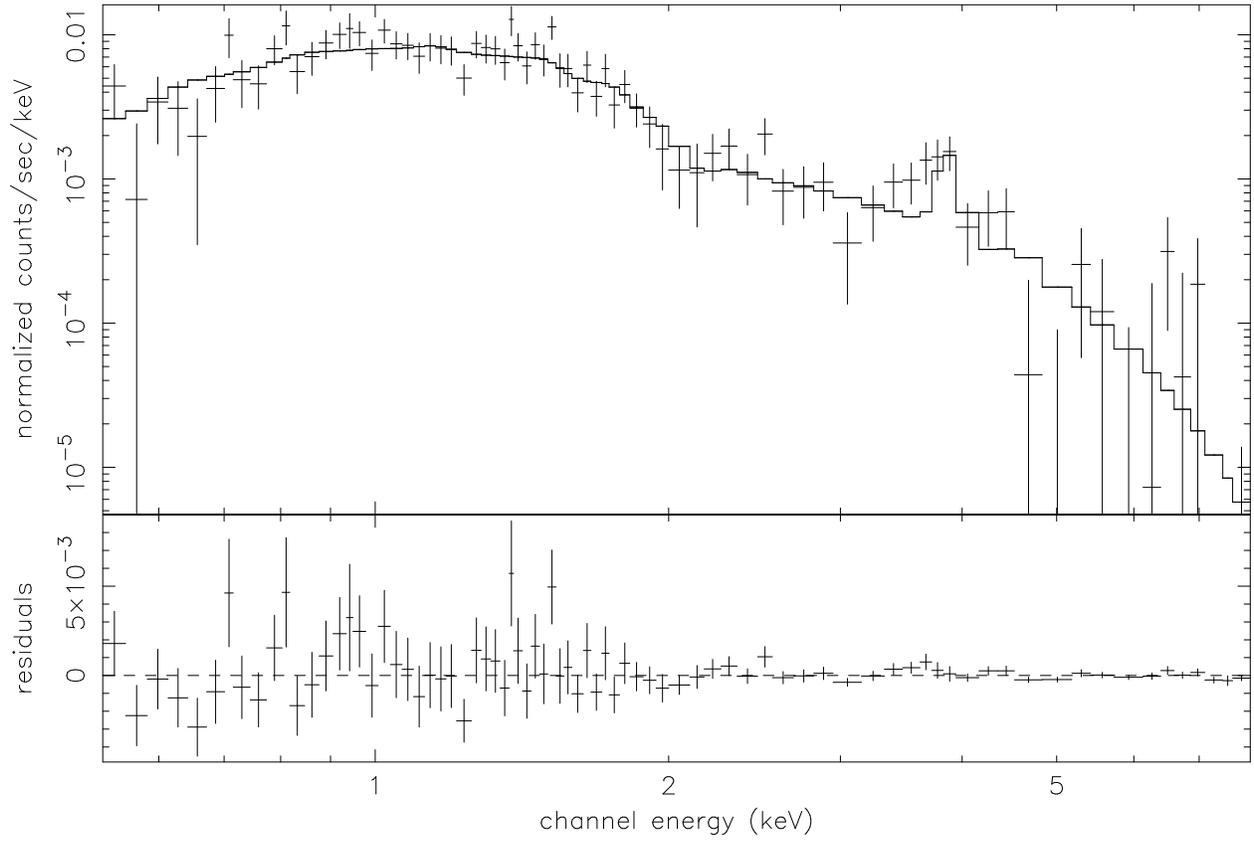}
\caption{\label{fig:j1113_spec}Spectrum and best-fit model of
ClJ1113.1$-$2615.}
\end{center}
\end{figure}

An initial radial profile suggested a spectral radius of $50{\arcs}$,
and so a circular region of this size, centred on the X-ray centroid ($\alpha=11^h13^m5\dotsec23$, $\delta=-26\degree15\arcmin41\dotarcs4$)
and excluding a region of radius $2{\arcs}$ around the point source to
the north-east, was used to produce a spectrum containing around 1000
net counts. The spectrum was grouped to obtain a minimum of 20 counts
per bin, and fit with an absorbed \MEKAL model (including the ACISABS absorption model discussed in Appendix \ref{app1}), with the absorbing
hydrogen column density frozen at the Galactic value of
$5.47\times10^{20}\pcmsq$ \citep{dic90}. The redshift was frozen at
$0.725$, and all fits were performed in the energy range of
$0.5-8\keV$. The spectrum and the best-fit model are shown in
Fig. \ref{fig:j1113_spec}.

\clearpage

\clearpage

The best fit to the spectrum was obtained for a temperature of
$4.3^{+0.5}_{-0.4}\keV$ and an abundance of $0.62^{+0.25}_{-0.22}$ solar,
with a reduced \chisq value of $1.00$ ($62$ degrees of freedom). We repeated the fit with the
abundance frozen at the canonical value of $0.3$, and found a
temperature of $4.4^{+0.7}_{-0.5}\keV$ and a reduced \chisq value of
$1.02$ ($63$ degrees of freedom). A redshifted iron line is prominent at $3.8\keV$. Thawing the
redshift parameter of the spectral model led to a best fit for a value
of $z=0.745^{+0.084}_{-0.015}$, which is higher than the optically
derived value of $z=0.725$ but not significantly so. Similar redshift
differences have been observed in the \Chandra spectra of other
clusters \citep[\egc][]{sch01a} and are likely to be due to
calibration errors. If so, such errors in the position of the spectral
lines will also affect the spectral fits of the metallicities. However, as the differences in the fit parameters caused by freezing, or fitting the metal abundance are well within their $1 \sigma$ errors, we use those values derived with the fitted metallicity in our further analysis.

A hardness ratio map of the cluster was produced by creating images in the energy bands $0.5-1.5\keV$ (soft band) and $1.5-8\keV$
(hard band), subtracting a constant background level in each
energy band, and dividing the hard image by the soft image. The images were both exposure corrected with the same, broad band exposure map. Strictly, they should be corrected by exposure maps generated for the same range of energies as the images, however this is a very small effect compared to the large errors due to counting statistics. The low
number of photons available forced a large pixel size to be used in
order to reach $\geq20$ counts per pixel, so the resolution of the
image was low ($110\kpc$/pixel). Spectra were simulated, using an absorbed \MEKAL model at the cluster redshift, for a series of temperatures, and the number of photons in the hard and soft bands defined above were divided to give a hardness ratio. In this way, a look-up table was constructed, that allowed the conversion of the measured hardness ratios into temperatures.
The limits of the data allow us only
to state that the cluster shows no significant departures from a temperature of $5\pm3\keV$
on scales of $\sim100\kpc$. In
particular, there is no evidence for a central cooling flow, although
the constraints are weak.

\subsection{Spatial results}\label{sect:j1113spatial}
The cluster X-ray emission was then examined spatially, as described
in section \ref{sect:spatial}. The source model included a constant
background flux of $3.95\times10^{-10}\thinspace
\mathrm{counts}\pcmsq\ps$ which was estimated from an independent 2D
fit to a background region of the same CCD. Four point sources in the
fitting regions were masked out. After convolution with an appropriate
PSF model and multiplication with an appropriate exposure map, the
model was fit to the data in Sherpa using a maximum-likelihood
algorithm and the C statistic. The best-fit model
parameters were $r_c=14.6^{+1.2}_{-2.2}{\arcs}$ ($105^{+9}_{-16}{\kpc}$),
$\beta=0.67^{+0.03}_{-0.05}$, and $e=0.2$. The background level was also allowed to fit, which gave no significant change to the best fitting parameters. Errors were computed with
all parameter values free to vary, but were not computed for the
eccentricity because of the computational load involved.

\begin{figure}
\begin{center}
\plotone{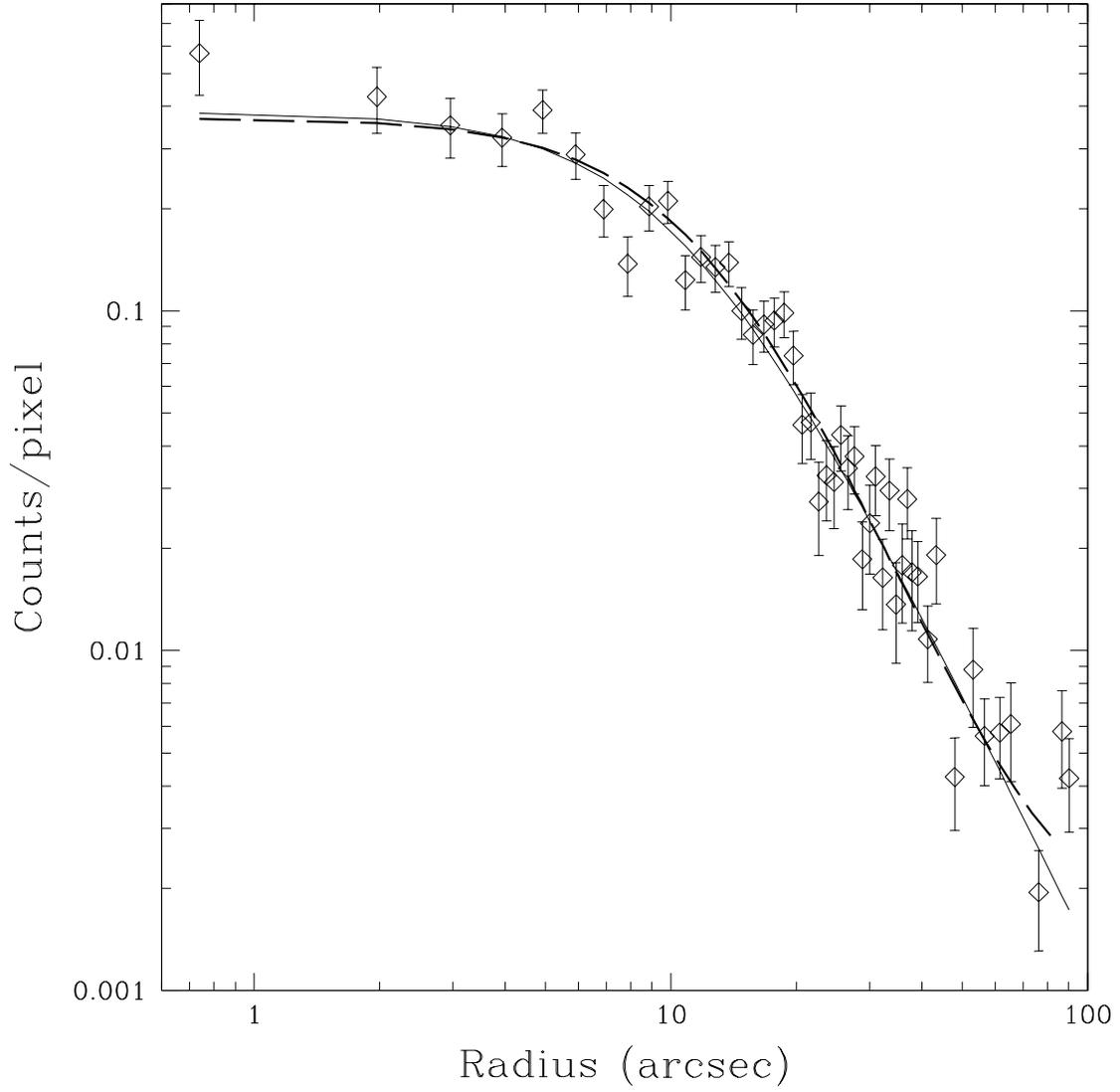}
\caption{\label{fig:j1113_radprof}Surface-brightness profile of
ClJ1113.1$-$2615. The solid line shows the best-fit 1D $\beta$-model,
and the dashed line shows a profile of the best-fitting 2D model.}
\end{center}
\end{figure}

The image was then divided by a normalised exposure map to produce a
exposure corrected image (with units of counts), and a radial profile was
produced excluding the same point sources, estimating the background
from the same region as the 2D fitting, and centred on the best-fit
position of the 2D model. The radial bins were adaptively sized so
that their minimal width was 2 pixels ($\approx 1\arcs$) and the
minimal snr value per bin was $3$. Emission was detected at this level
out to a radius of $92.0\arcs$, which is similar to the size of the
region within which the 2D fitting was performed. The best-fit
parameters of a $\beta$-profile fit to these data are
$r_c=10.5^{+1.3}_{-1.2}{\arcs}$ ($76^{+9}_{-9}{\kpc}$) and $\beta=0.58^{+0.03}_{-0.03}$, with
a reduced $\chisq$ value of $1.6$ ($46$ degrees of freedom). These fit results do not agree well
with those of the 2D fit, however, it is easily shown that the
disagreement is due to the eccentricity of the cluster emission. An
image of the best-fitting 2D model, convolved with the PSF and
multiplied with the exposure map, was used to produce a radial profile
in exactly the same manner as above. We obtained best-fit values of
$r_c=11.2^{+1.2}_{-1.1}{\arcs}$ ($81^{+9}_{-9}{\kpc}$) and $\beta=0.60^{+0.03}_{-0.03}$, in
good agreement with those found for the radial profile of the real
data. We thus consider the most reliable measurements to be those from
the 2D fit. The radial profile, best-fitting $\beta$-model, and
profile of the best-fitting 2D model are shown in
Fig. \ref{fig:j1113_radprof}.

The reduced $\chisq$ statistic of $1.6$ indicates that the 1D fit is
not a very good one, however, given that we are fitting a circular
model to an elliptical cluster, and that the profile of the 2D model
appears to agree well with the data, we believe that the 2D model is a
good description of the data.

\clearpage

\clearpage

To better examine any discrepancies between the data and our
$\beta$-model, we subtracted the best-fitting 2D model from the data,
and smoothed the resulting image with a Gaussian kernel of fixed size
$\sigma=3{\arcs}$, thus producing a residual map. The residuals are
evenly distributed with no significant features, again indicating that
the 2D fit provides a good description of the data.

Using Eqn.~\ref{virrad} we estimate the virial radius of the cluster
to be $1.3\pm0.1\Mpc$. Our 1D analysis of the X-ray surface
brightness detects cluster emission out to approximately $51\%$ of
this value, whereas the spectral radius, within which the immediately
detected flux is measured, corresponds to only $\approx27\%$ of the
virial radius. In order to estimate the total flux, and hence the
total luminosity of the cluster, the observed flux was extrapolated
out to the virial radius using the best-fit $\beta$-model. To allow
comparisons with other work, this extrapolation was carried out in
both of our chosen cosmologies (the value of $r_v$ is cosmology
dependent, as is the conversion between the angular and physical size
of the spectral radius). We found the scaling factor to be
$\approx1.3$ for either cosmology. The bolometric, unabsorbed, total
cluster flux is thus $F(r_v)=1.38^{+0.13}_{-0.12}\times10^{-13}\flux$,
corresponding to a bolometric luminosity of $L(r_v)=3.3^{+0.3}_{-0.3}
\times10^{44} \ergps$ (\LCDM ~cosmology).

In order to allow a comparison of the result from the flux measurement
with \Chandra with those of previous observations, we also
extrapolated to the virial radius the unabsorbed flux in the
$0.5-2\keV$ band using our alternative Einstein-de Sitter
cosmology. We find $F(r_v)=5.7\pm0.5\times10^{-14}\flux$,
which is marginally lower than the value of $7.4\pm1.1\times10^{-14}\flux$
measured with \ROSAT \citep{per02}. This lower flux can be explained 
by the exclusion of point sources in this analysis that were
not resolved by \ROSAT. \Chandra resolves $5$ point sources in the region where the \ROSAT flux was measured. The fluxes of these sources were estimated, assuming the best fitting spectral model to the brightest point source within $2\Mpc$ of the cluster. This was a absorbed power law, with a photon index $\gamma=2.09\pm0.05$. The combined $0.5-2\keV$ flux in these five point sources was $1.3\times10^{-14}\flux$, which brings the \Chandra flux up to $F(r_v)=7.0\pm0.5\times10^{-14}\flux$, in good agreement with the \ROSAT flux.

For further comparison with other work (section
\ref{sect:discussion}), the bolometric luminosity was also computed
for an Einstein-de Sitter cosmology, giving a value of
$3.9^{+0.4}_{-0.3}\times10^{44}\ergps$.

The central hydrogen number density in ClJ1113.1$-$2615 was found to
be $n_H=7.3\pm0.6\times10^{-3}\pcc$, giving a gas mass within
the spectral radius, and extrapolated out to the virial radius, of
$M_g(r_s)=5.7^{+0.2}_{-0.2}\times10^{12}M_{\odot}$ and
$M_g(r_v)=3.0^{+0.4}_{-0.4}\times10^{13}M_{\odot}$, respectively. The
total gravitating mass of the cluster is estimated at
$M(r_s)=1.1^{+1.4}_{-1.2}\times10^{13}M_{\odot}$ within the spectral
radius of ClJ1113.1$-$2615; extrapolating to the virial radius we find
$M(r_v)=4.3\pm0.1\times10^{14}M_{\odot}$. This gives a gas
mass fraction $(M_g/M)$ within the spectral radius of
$0.05\pm0.01$, and of $0.07\pm0.01$ within the
virial radius. A comparison of these cluster properties with those
measured for a sample of local groups and clusters is performed in
section \ref{sect:discussion}.

\section{Cluster ClJ0152.7$-$1357} \label{sect:J0152}
\begin{figure*}
\begin{center}
\plotone{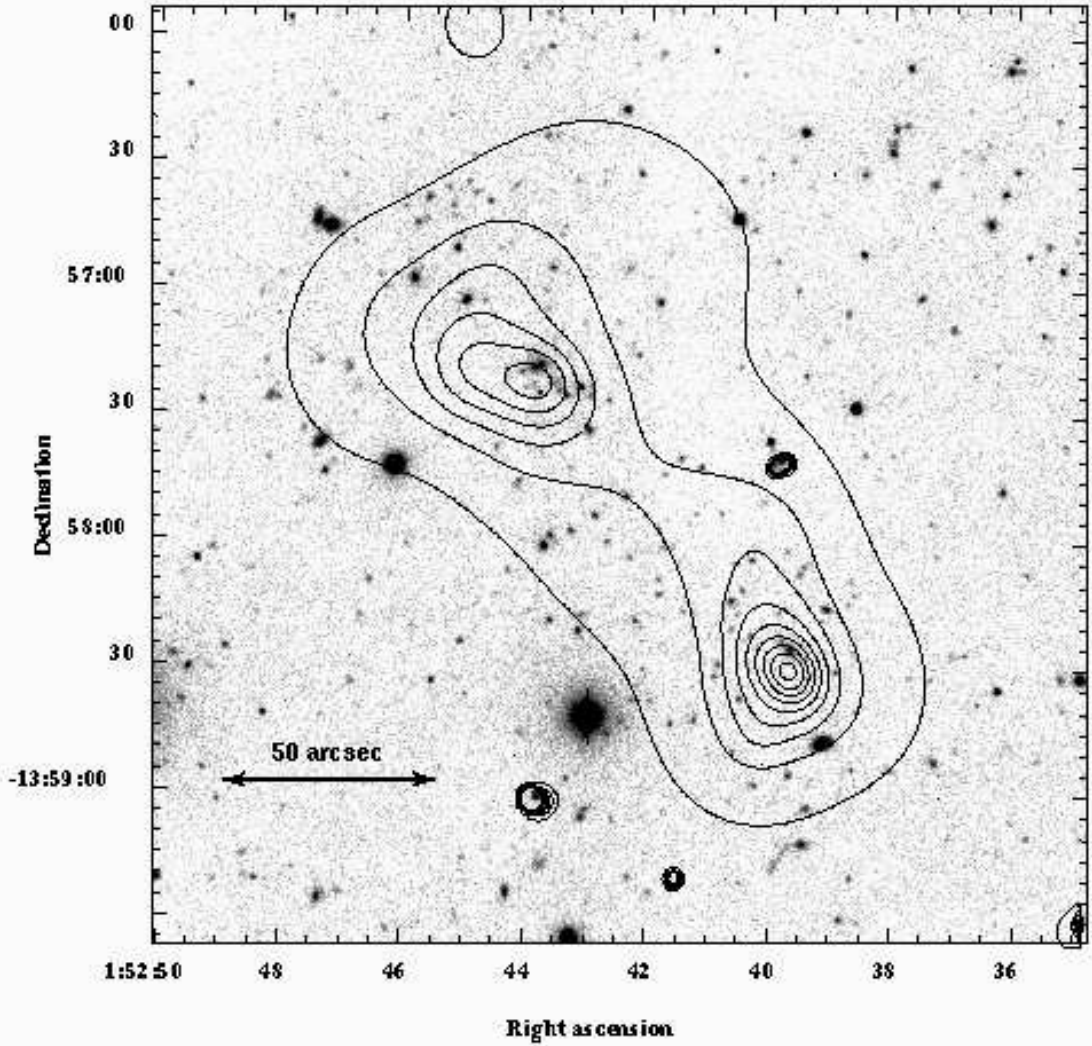}
\caption{\label{fig:j0152_overlay}Adaptively smoothed \Chandra X-ray
contours overlaid on a Keck II I-band image of ClJ0152.7$-$1357. The
contours were produced from an exposure corrected image in the energy
band $0.5-5\keV$, which was adaptively smoothed so that all features
were significant at the 99\% level. The contours start at
$0.03\mathrm{~counts ~pixel^{-1}}$, and are logarithmically
spaced, by a factor of $1.5$.}
\end{center}
\end{figure*}

Upon its discovery in archival \ROSAT PSPC data, ClJ0152.7$-$1357 was
the most X-ray luminous distant ($z>0.8$) cluster known
\citep{ebe00}. Its X-ray morphology as seen with \ROSAT strongly
suggested complex substructure, possibly the result of a merger in
progress. At the cluster redshift of $z=0.833$ 1 arcsecond corresponds
to $7.6\kpc$ (\LCDM). Fig. \ref{fig:j0152_overlay} shows \Chandra X-ray
contours overlaid on a Keck II (LRIS) I-band image of the cluster.
 The contours clearly show two peaks of X-ray
emission, coincident with two concentrations of galaxies. It appears
that the contours are more tightly grouped along the south-west edge
of the northern subcluster, and more extended to the north-east. This
is suggestive of the northern subcluster moving toward the
southern cluster component.

It is interesting to note that the inner contours of the southern
subcluster are displaced to the south of the subcluster galaxies, with
the local X-ray centroid $\sim5{\arcs}$ south of the galaxies. The
astrometry of both the optical and X-ray data (using X-ray point
sources in the field of view) were checked against the Automatic Plate
Measuring (APM) source catalogue, and found to be accurate to within
$\sim0.5{\arcs}$; hence the displacement appears to be real. Similar
displacements have been observed in other merging galaxy clusters
\citep[\egc][]{mar02}, and can be explained by the collisionless
galaxies (and presumably dark matter) moving ahead of the ICM which is
slowed by ram pressure.
 
The X-ray centroids of the two subclusters (North: $\alpha=1^h52^m44\dotsec18$, $\delta=-13\degree57\arcm15\dotarcs84$; South: $\alpha=1^h52^m39\dotsec89$, $\delta=-13\degree58\arcm27\dotarcs48$) are $95{\arcs}$ apart,
equivalent to $722\kpc$ in our \LCDM cosmology. The question of
whether the two subclusters are gravitationally bound is addressed
later in this section.

There is also a region of extended low surface brightness emission to the east of the cluster ($\alpha=1^h52^m52\dotsec42$, $\delta=-13\degree58\arcm5\dotarcs52$) associated with an overdensity of galaxies. Three galaxy redshifts confirm the existence of a group of galaxies at $z=0.844\pm0.002$ \citep{dem02}. The limited statistics of both the optical and X-ray data prohibit detailed investigation of this system, but it would appear to be an infalling galaxy group. Note that this feature is not visible within the field of Fig. \ref{fig:j0152_overlay}.

\clearpage

\clearpage

\begin{figure}
\begin{center}
\plotone{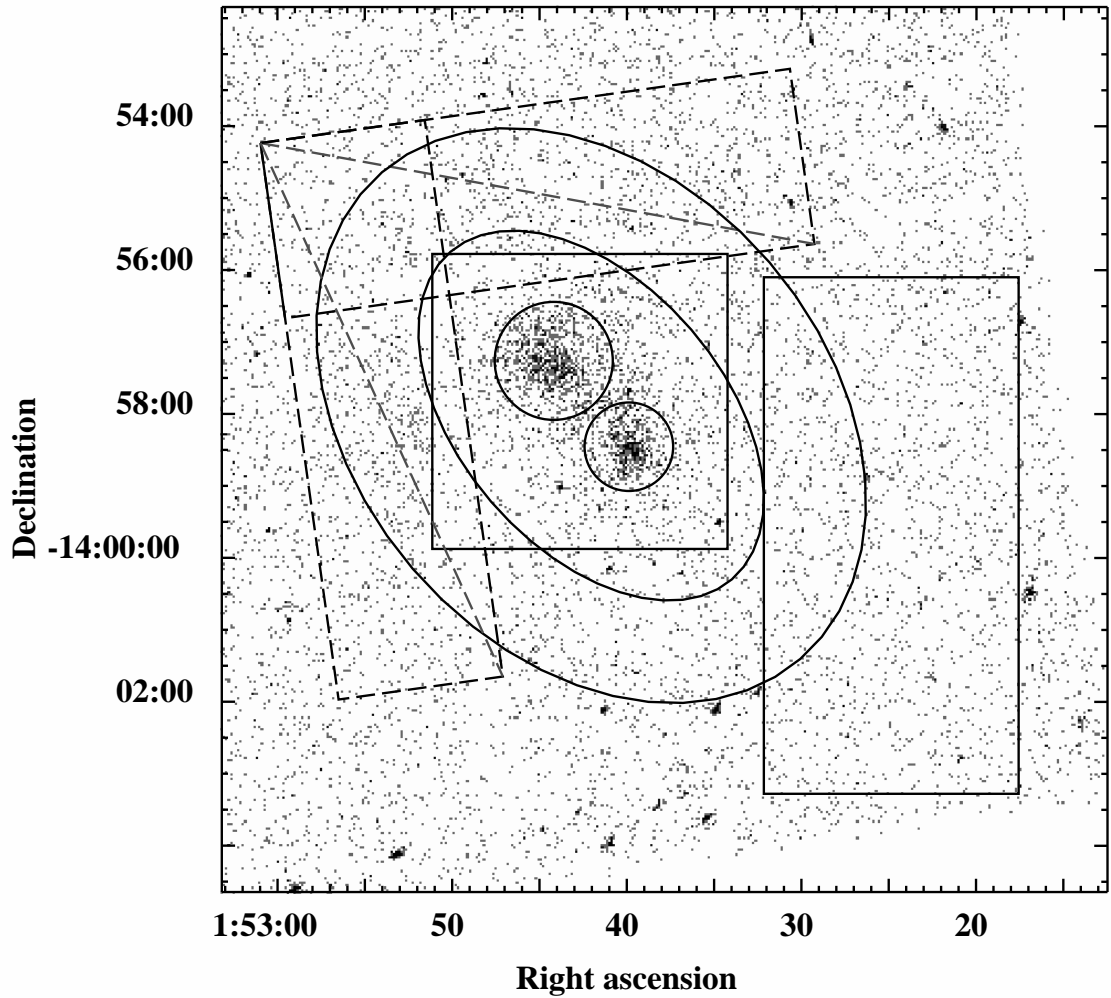}
\caption{\label{fig:j0152regions}Chandra X-ray image of ClJ0152.7$-$1357 with pixels of $2\arcs$. The regions used for imaging and spectral analysis are shown (see text).}
\end{center}
\end{figure}

The regions used for the extraction of different products are shown in Fig. \ref{fig:j0152regions}, overlayed on a binned X-ray image. The circular regions were used to extract the source spectra, with the elliptical annulus as a background (excluding the dashed, barred rectangular regions). The non-barred rectangular regions were used for the imaging analysis. Point sources were excluded from all spectral and imaging analysis.

\clearpage

\clearpage

\subsection{Spectral results}
\begin{figure}
\begin{center}
\plotone{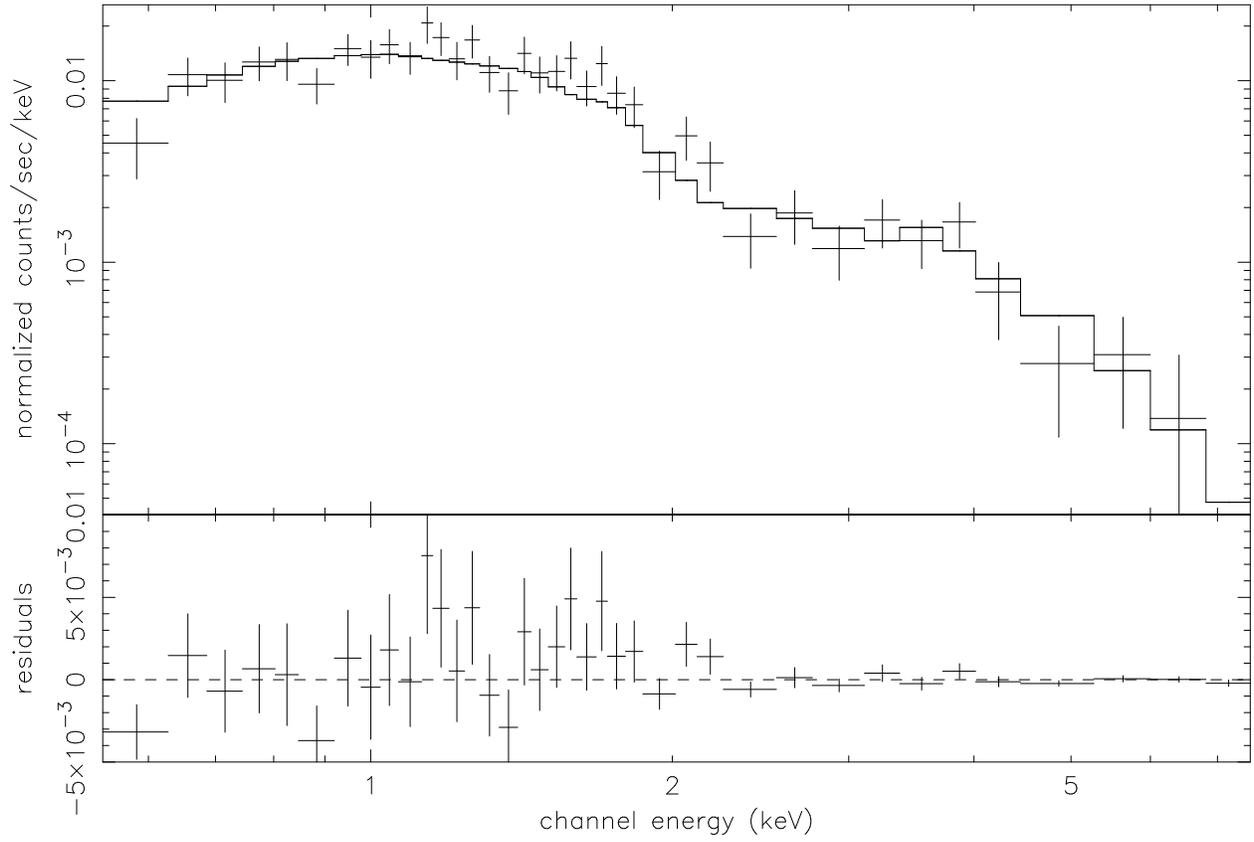}
\caption{\label{fig:j0152north_spec}Spectrum and best-fit model of the
northern subcluster of ClJ0152.7$-$1357.}
\end{center}
\end{figure}

\begin{figure}
\begin{center}
\plotone{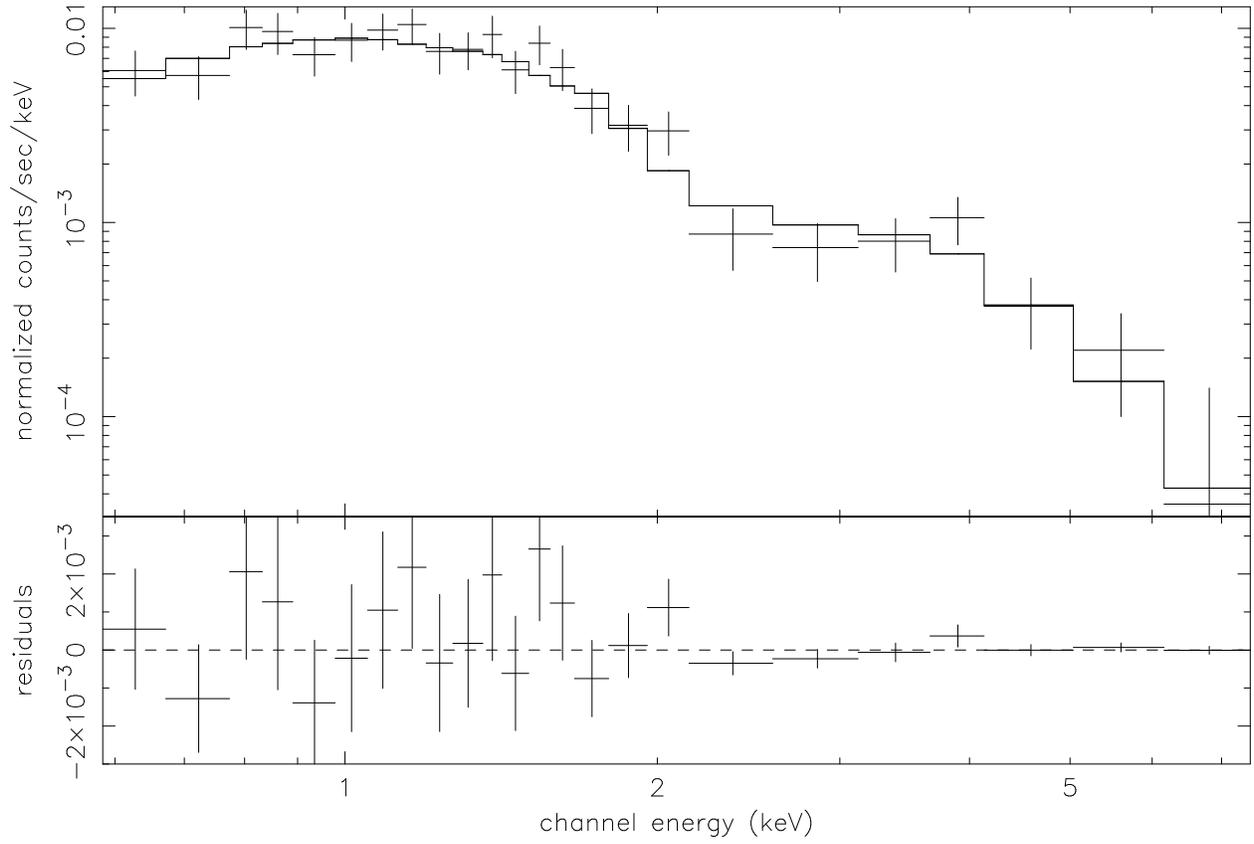}
\caption{\label{fig:j0152south_spec}Spectrum and best-fit model of the
southern subcluster of ClJ0152.7$-$1357.}
\end{center}
\end{figure}

Spectra were extracted within circular regions of radius $49{\arcs}$
and $37{\arcs}$ of the centroids of the northern and southern
subcluster respectively, and were fit with an absorbed \MEKAL
model (again including the ACISABS absorption model discussed in Appendix \ref{app1}). The northern cluster region contained around $750$ net counts,
and the southern cluster around $450$. The absorbing hydrogen column
density was frozen at the Galactic value \citep{dic90} for the
position of each of the subcluster centroids
(north:$1.55\times10^{20}cm^{-2}$,
south:$1.54\times10^{20}cm^{-2}$). The \MEKAL redshift was frozen at
$0.833$ for both cluster components, and the spectra were fit in the
$0.5-8\keV$ energy range, after having being grouped into bins
containing $\geq20$ counts each. The spectra and best-fit models are
shown in Figs. \ref{fig:j0152north_spec} and
\ref{fig:j0152south_spec}.

\clearpage

\clearpage

The spectral fits yielded a temperature of $5.6^{+1.0}_{-0.8}\keV$,
and an abundance of $0.15^{+0.22}_{-0.15}$ solar, with a reduced \chisq
value of $0.9$ ($36$ degrees of freedom) for the northern subcluster. The southern subcluster
was found to be slightly, but not significantly, cooler at a
temperature of $4.8^{+1.4}_{-0.9}\keV$ with a metal abundance of
$0.70^{+0.72}_{-0.44}$ solar; the reduced \chisq was $0.7$ ($35$ degrees of freedom). We note that
the abundances are poorly constrained and so we prefer to fit
for the temperature with the abundance frozen at $0.3$ solar. This
then yields best-fit values of $5.5^{+0.9}_{-0.7}\keV$ and
$5.2^{+1.1}_{-0.9}\keV$ for the northern and southern subclusters
respectively, without significant changes to the reduced \chisq
values.

\ROSAT observations of this system gave a temperature estimate of
$5.9^{+4.4}_{-2.1}\keV$ \citep{ebe00}, while an analysis of a
\BeppoSAX observation produced a temperature of
$6.5^{+1.7}_{-1.2}\keV$ and a metallicity of $0.53^{+0.29}_{-0.24}$
solar \citep{del00}. Finally, \citet{joy01} use Sunyaev-Zel'dovich
effect imaging of the cluster to estimate its ICM temperature at
$8.5^{+2.0}_{-1.5}\keV$. While these quantities were all measured for
the cluster as a whole, the X-ray temperatures are consistent with that of
either subcluster as measured with \Chandra. The \BeppoSAX metallicity
is in good agreement with that of the southern cluster and consistent
with that of the northern subcluster, which is not surprising given
the size of the $1\sigma$ errors of the measurements. The Sunyaev-Zel'dovich temperature estimate is higher than that of the \Chandra temperature measurement, though at less than $2 \sigma$.

\subsection{Spatial results}\label{sect:j0152spatial}
The spatial distribution of the X-ray emission of ClJ0152.7$-$1357 was
examined as described in section \ref{sect:spatial}. The system was
modelled with two 2D $\beta$-profiles (one for each subcluster) and a
constant background flux of $3.37\times10^{-10}\thinspace
\mathrm{counts}\pcmsq\ps$ as estimated from an independent 2D fit to a
background region of the same CCD. Five point sources in the fitting
regions were masked out, and the model was fit to the data in Sherpa,
using a maximum-likelihood algorithm and the C statistic. The
best-fit parameter values for the northern subcluster were
$r_c=32.7^{+8.3}_{-4.3}{\arcs}$ ($248^{+63}_{-33}{\kpc}$), $\beta=0.73^{+0.13}_{-0.06}$, and
$e=0.08$. The fit to the southern subcluster yielded
$r_c=16.1^{+3.7}_{-2.6}{\arcs}$ ($122^{+28}_{-20}{\kpc}$), $\beta=0.66^{+0.08}_{-0.06}$, and
$e=0.00$. Errors were computed with all parameters free to vary.

\begin{figure}
\begin{center}
\plotone{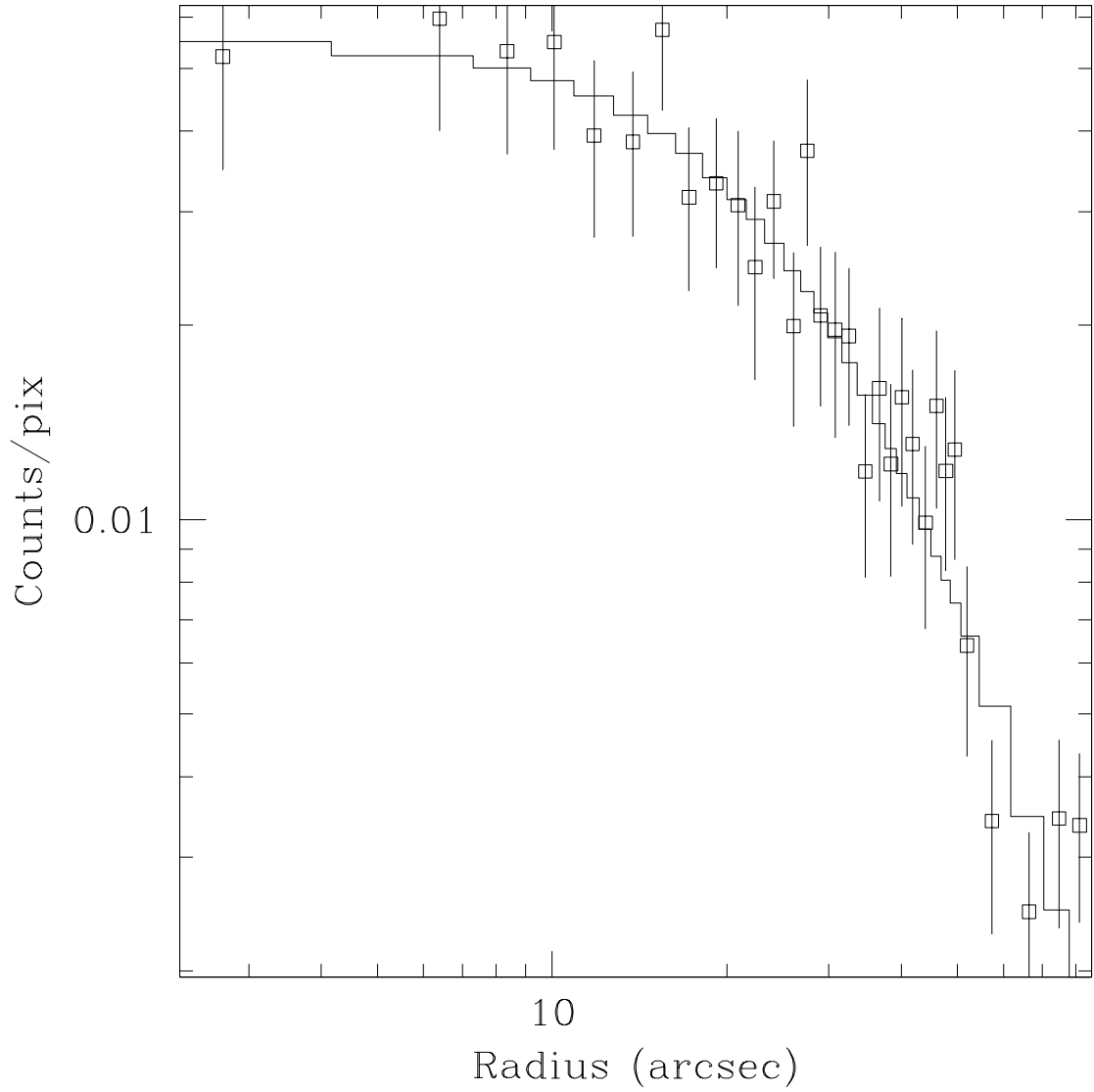}
\caption{\label{fig:j0152Nprof}X-ray surface brightness profile of the
northern subcluster of ClJ0152.7$-$1357 in a semi-circular region
spanning $180^{\circ}$ away from the direction of the merger. The line
shows the best-fit 1D $\beta$-model.}
\end{center}
\end{figure}

\begin{figure}
\begin{center}
\plotone{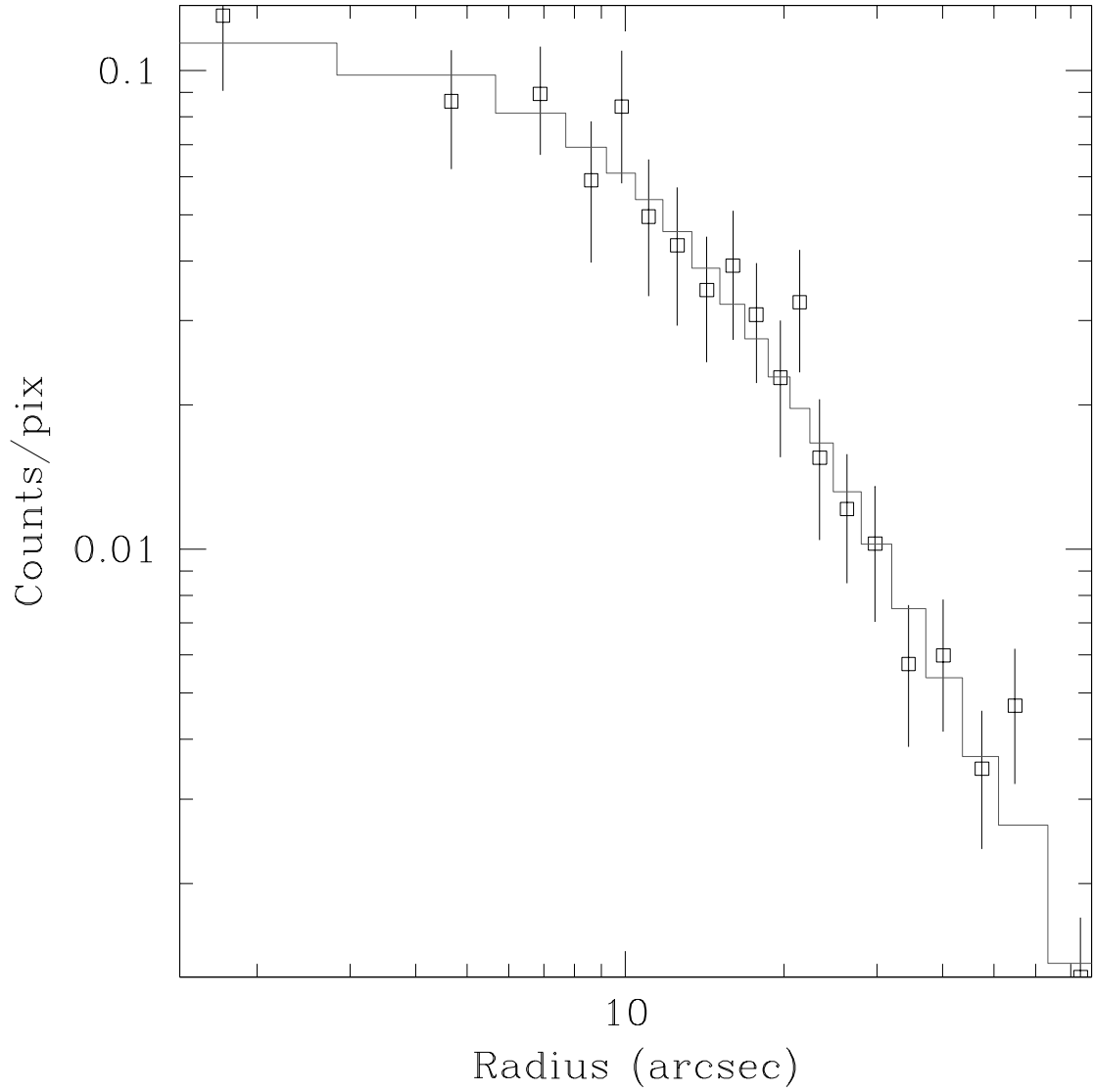}
\caption{\label{fig:j0152Sprof}X-ray surface brightness profile of the
southern subcluster of ClJ0152.7$-$1357 in a semi-circular region
spanning $180^{\circ}$ away from the direction of the merger. The line
shows the best-fit 1D $\beta$-model.}
\end{center}
\end{figure}

Radial profiles of the X-ray surface brightness were then produced for
each subcluster. In both cases the profiles were extracted from a
semi-circular region spanning $180^{\circ}$ away from the direction of the
merger to avoid contamination by the emission from the other
subcluster. The radial bins were adaptively sized so that their
minimum width was 2 pixels ($\approx 1\arcs$) and the minimal snr
value per bin was $3$. In the case of the northern subcluster,
emission was detected out to $85\arcs$ at this level; in the case of
the southern subcluster, the detection extended to $90\arcs$. These
radii are in good agreement with the size of the 2D fitting
region. For the northern subcluster a best-fitting radial profile was
obtained for $r_c=33^{+12}_{-8}{\arcs}$ ($251^{+91}_{-61}{\kpc}$) and
$\beta=0.74^{+0.23}_{-0.13}$, with a reduced $\chisq$ value of
$0.7$ ($28$ degrees of freedom). This fit agrees well with the one derived from the best-fitting
2D model; the data and best-fitting 1D model are shown in
Fig. \ref{fig:j0152Nprof}. The best-fitting model to the southern
profile is parameterised by $r_c=11.6^{+3.2}_{-2.6}{\arcs}$ ($88^{+24}_{-20}{\kpc}$) and
$\beta=0.57^{+0.07}_{-0.05}$, with a reduced $\chisq$ value of $0.4$ ($17$ degrees of freedom);
data and model are shown in Fig. \ref{fig:j0152Sprof}. These values
are lower than those found with the 2D fitting, but not significantly
so. In only considering half of the emission from each subcluster the
errors, already large due to poor photon statistics, are increased
further, leading to yet greater uncertainties in the 1D fit results
and low \chisq values. 2D fitting seems to be the best approach to
model this complex system as the emission from both subclusters can
be fit simultaneously, thus making the most of the available
photons. This method also has the advantage that any excess emission
over that expected from two clusters overlapping in projection will be
apparent in the residuals of the fit.

\clearpage

\clearpage

\begin{figure}
\begin{center}
\plotone{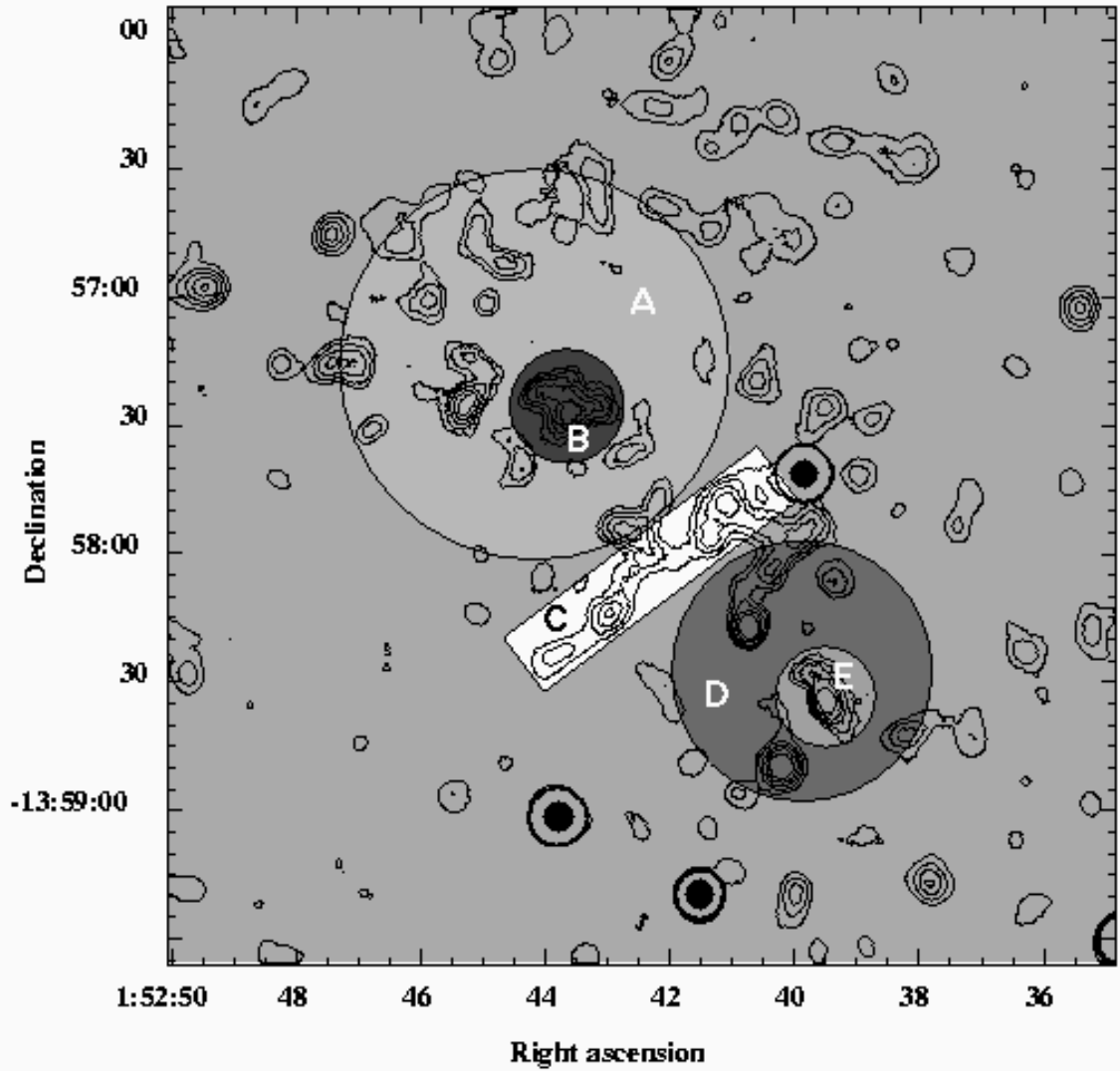}
\caption{\label{fig:HS}Hardness ratio map of ClJ0152.7$-$1357, overlayed with contours of excess X-ray emission above the best-fitting surface brightness model. Lighter colours represent harder emission; the values for each region are given in Table \ref{tab:HSvalues}. Note that the three point sources which appear black were excluded from this analysis}
\end{center}
\end{figure}

A residual map was produced by subtracting the best-fit model
(convolved with the PSF and multiplied by the exposure map) from the
data, and then smoothing the resulting image with a Gaussian kernel of
size $\sigma=3{\arcs}$. Contours of the positive residuals are shown
in Fig. \ref{fig:HS}, overlaid on a hardness ratio map of the cluster emission, produced using the regions of excess emission as a guide. The hardness ratios of the emission were calculated as described in section \ref{sect:j1113spectral}. 
Immediately apparent is a
linear feature of excess emission between the two subclusters,
oriented perpendicular to their merger axis. This narrow feature
(width $<5\arcs$) is suggestive of a shock front between the merging
subclusters. It is, however, significant at only the $4.2\sigma$
level, in a region carefully chosen to maximise its significance (the
rectangle in Fig. \ref{fig:HS}; note that this
region excludes a point source at the western end of the feature). This region also has harder emission than either of the subclusters, corresponding to hotter gas (see Table 1), 
as one would expect in a region of compression, though this result is not strongly significant. If
confirmed, this feature would constitute the first direct detection of
a shock in the early stages of a cluster merger. No other systematic
residuals are apparent in the outer cluster regions, indicating that,
overall, our model fits the data well.

Additional residuals of excess emission are, however, observed in the
core of each subcluster (regions C and F in
Fig. \ref{fig:HS}), suggesting the possible presence of cooling
flows \citep[\egc][]{fab94b}, but these residuals are significant at only the
$3\sigma$ level. The core of the northern subcluster may have the softest emission in this system (but the errors are large), 
consistent with a cooling flow, but the opposite is the case in the southern
subcluster.  Simulations predict that cooling flows are disrupted
in the late stages of major mergers such as this one \citep{rit02}.
The possible existence of at least one cool cluster core in this developing merger
suggests that the central regions of the approaching systems are still
fairly undisturbed and likely close to hydrostatic equilibrium.


\clearpage

\begin{deluxetable}{ccc}
\tablecaption{\label{tab:HSvalues}Values of the hardness ratios, and corresponding temperatures measured for the regions shown in Fig. \ref{fig:HS}}
\tablewidth{0pt}
\tablehead{
\colhead{Region} & \colhead{Hardness ratio} & \colhead{Temperature ($\keV$)}}
\startdata
A & $0.82\pm0.08$ & $8.8^{+2.6}_{-2.0}$ \\
B & $0.58\pm0.11$ & $4.4^{+1.5}_{-1.1}$ \\
C & $1.24\pm0.29$ & $>14.1$ \\
D & $0.68\pm0.10$ & $5.7^{+2.0}_{-1.3}$ \\
E & $0.78\pm0.14$ & $7.7^{+4.8}_{-2.6}$ \\ 
\enddata
\end{deluxetable}

\clearpage

\clearpage

Using Eqn.~\ref{virrad} the virial radius of the northern subcluster
is estimated to be $1.4\pm0.1\Mpc$. The spectral radius used
in our analysis thus corresponds to approximately $0.27\,r_v$. For the
southern cluster we find our spectral radius to reach about $20\%$ of
the virial radius of $r_v=1.4\pm0.1\Mpc$.

Cluster fluxes for the two cluster components were initially measured
within the respective spectral radii, and then, as before, scaled out
to the virial radius by integrating the respective best-fit surface
brightness model to yield approximate total fluxes. The unabsorbed
total fluxes were used to estimate the luminosities of the
subclusters. For the northern subcluster we find a bolometric
luminosity of $L(r_v)=1.0\pm0.2\times10^{45}\ergps$; for the
southern subcluster the corresponding value is
$L(r_v)=5.8^{+1.2}_{-0.9}\times10^{44}\ergps$. Combined, these values
lead to a total bolometric luminosity of ClJ0152.7$-$1357 of
$L(r_v)=1.6\pm0.2\times10^{45}\ergps$.

The subcluster fluxes measured in the $0.5-2\keV$ band, again
extrapolated to $r_v$, were combined to estimate the total flux of
ClJ0152.7$-$1357 in this band for comparison with earlier
measurements, adopting the Einstein-de Sitter cosmology. We find
$F(r_v)=1.83^{+0.22}_{-0.18}\times10^{-13}\flux$, slightly lower than
the value of $2.9^{+0.18}_{-0.18}\times10^{-13}\flux$ measured from
the \ROSAT data \citep{ebe00}. The higher value found in the \ROSAT
data could be explained by a contribution from point sources which
were unresolved by that instrument, but which are excluded here, and
by the fact that the \Chandra flux is extrapolated out to the virial
radius, rather than to infinity. For completeness' sake we also
computed the bolometric luminosity of each subcluster in our
alternative, Einstein-de Sitter cosmology (see Table
\ref{tab:summaryH50}). The total bolometric luminosity of
ClJ0152.7$-$1357 (both subclusters combined) is
$L(r_v)=1.81^{+0.21}_{-0.18}\times10^{45}\ergps$ in this cosmology, in
good agreement with the value of
$2.2^{+0.5}_{-0.5}\times10^{45}\ergps$ derived from the \BeppoSAX
observation \citep{del00}.

Following the procedure detailed in section \ref{sect:dataprep}, the
central gas densities, gas masses, total gravitating masses, and gas
mass fractions of each subcluster were computed, with results as
listed in Table \ref{tab:summary} and Table \ref{tab:summaryH50} for
the two different cosmologies. We confirm that ClJ0152.7$-$1357 is
indeed a massive cluster; the combined total mass of the two
subclusters extrapolated to the virial radius is
$1.1\pm0.2\times 10^{15}\Msol$ (\LCDM cosmology).

The Sunyaev-Zel'dovich effect (SZE) work of \citet{joy01} quotes an
estimate of the mass of ClJ0152.7$-$1357 of
$\ga2\times10^{14}h_{100}^{-1}M_\odot$, within a single radius of
$65{\arcs}$ for the entire system. While the different geometry used
by \citet{joy01} prohibits a direct comparison of their estimate with
the masses derived here, we note that the masses of the subclusters
within the spectral radii ($\sim50\arcs$) combine to give
$2.4^{+0.4}_{-0.3}\times10^{14}h_{70}^{-1}M_\odot$ in good agreement
with the SZE estimate.

\subsection{Cluster dynamics}\label{sect:j0152dynamics}
We now attempt to address the question of whether ClJ0152.7$-$1357 is
a gravitationally bound system. The redshifts of nine member galaxies in each subcluster have been measured \citep{dem02}, allowing us to ascribe mean
redshifts of $z=0.8352\pm0.0017$ and $z=0.8294\pm0.0013$ to the northern and southern subclusters respectively. Via the relativistic redshift equation
\begin{eqnarray}
\frac{v}{c} & = & \frac{(z+1)^2-1}{(z+1)^2+1}
\end{eqnarray}
the difference between these redshifts can be translated into a
line-of-sight velocity difference between the subclusters of
$\Delta{v}=660\pm258\kmps$. This non-vanishing value of $\Delta{v}$
can be caused by two effects:
\begin{itemize}
\item The subclusters are separated by the difference in the
(line-of-sight) distances implied by their redshifts, i.e. $58\Mpc$,
and are being carried apart by the differential of the Hubble flow at
these distances.
\item The subclusters are at approximately the same distance, and
$\Delta{v}$ is a peculiar velocity reflecting the merger process.
\end{itemize} 
The former case implies a chance superposition of two separate
clusters, which can be regarded as extremely unlikely because the high
mass and temperature of the two subclusters makes them very rare
objects indeed. In the following we assess the probability of the
second scenario, which implies that the system is gravitationally
bound, using the system's observed properties and the technique
described by \citet{hug95}.

Given a line-of-sight velocity difference $v_r$ and a projected
separation $R_p$ between the two subclusters of ClJ0152.7$-$1357, the
true peculiar velocity $v$ and separation $R$ can be written as
\begin{eqnarray}
v_r & = & v\cos\Psi_v \\
R_p & = & R\sin\Psi_R,
\end{eqnarray}
where $\Psi_v$ and $\Psi_R$ are the inclination angles of the velocity
and separation vectors, respectively. Note that different values of
$\Psi_v$ and $\Psi_R$ imply an orbital component in addition to the
radial component of the relative motion of the subclusters. The system
is bound when the kinetic energy of the subclusters is less than the
gravitational potential of the system, a condition that can be
expressed as
\begin{eqnarray}
\label{eqn:bound}
v_r^2 - \frac{2GM}{R_p}\sin\Psi_R\cos^2\Psi_v & < & 0,
\end{eqnarray}
where $M$ is the total mass of the system. The values of the relevant
quantities, for the ClJ0152.7$-$1357 system, are $R_p=722\kpc$,
$v_r=660\kmps$ and
$M=1.13\times10^{15}M_{\odot}$. 

\clearpage

\begin{figure}
\begin{center}
\plotone{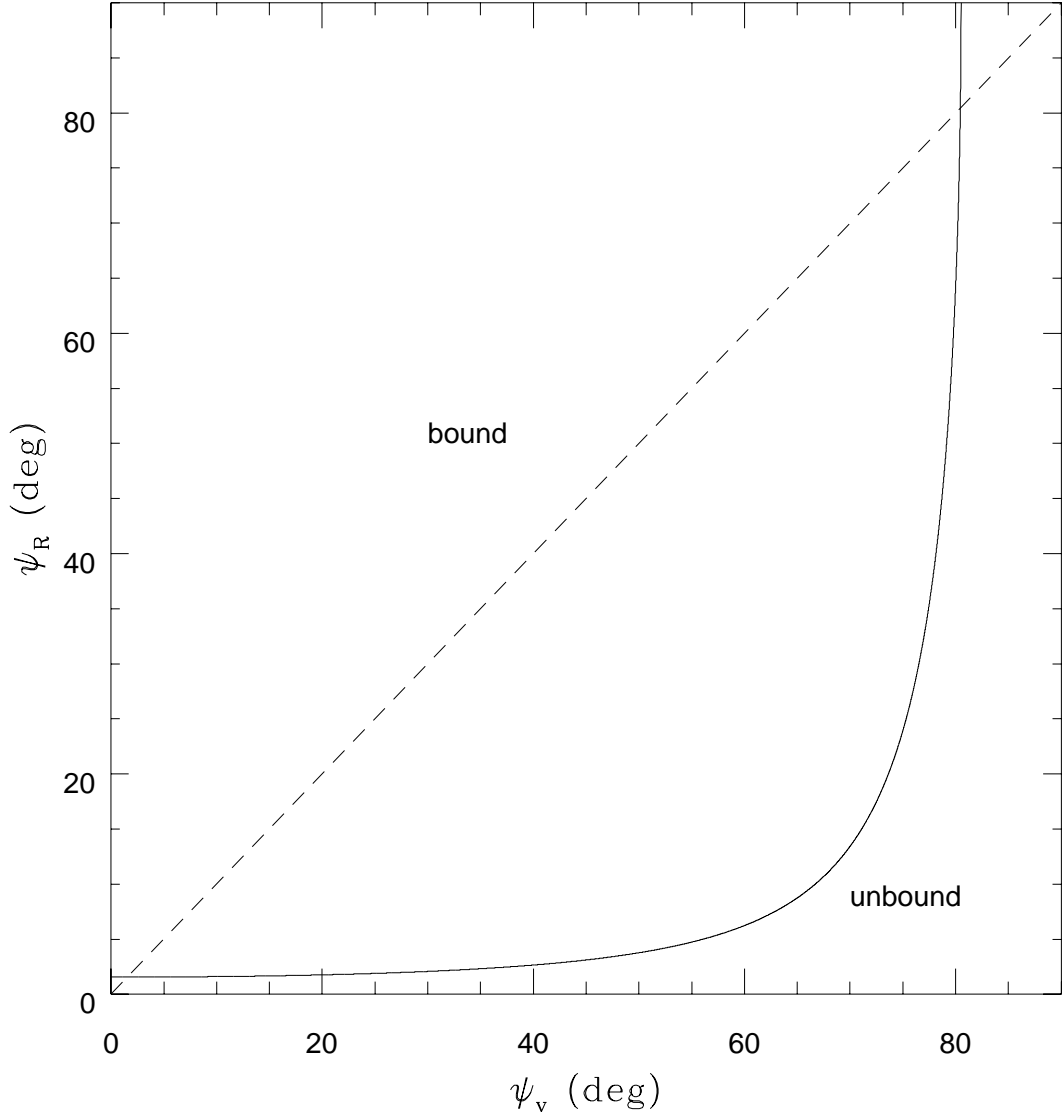}
\caption{\label{fig:boundplane}Regions of the $(\Psi_v,\Psi_R)$ plane
corresponding to the bound and unbound solutions of
Eqn. \ref{eqn:bound} for the ClJ0152.7$-$1357 system. The dashed line
is equivalent to a purely radial motion of the subclusters.}
\end{center}
\end{figure}

\clearpage

Fig. \ref{fig:boundplane} shows the regions of the $(\Psi_v,\Psi_R)$
plane corresponding to the bound and unbound solutions of
Eqn. \ref{eqn:bound}. The likelihood that the system is bound equals
the probability of the projection angles having the appropriate values
to place the system in the bound region of the $(\Psi_v,\Psi_R)$
plane. This probability is given by the ratio of the solid angle
through which a given projection angle is observable, to the solid
angle of a sphere. Hence the probability that
$\Psi_R$ has a value in the range $\Psi_R$ to $\Psi_R + \Delta\Psi_R$
is given by
\begin{eqnarray}
P(\Psi_R) & = & \int_{\Psi_R}^{\Psi_R+\Delta\Psi_R}\frac{2\pi\cos\Psi_R}{4\pi}d\Psi_R.
\end{eqnarray}
The overall probability that the system is bound can then be obtained
as the sum of the probabilities of the different combinations of
projection angles in the bound region of the $(\Psi_v,\Psi_R)$
plane. Without constraints on the range of permissible angles, we find
the probability that the system ClJ0152.7$-$1357 is gravitationally
bound, to be $0.78$.

\section{Discussion}\label{sect:discussion}
The measured and derived properties of ClJ1113.1$-$2615 and
ClJ0152.7$-$1357 are summarised in Table \ref{tab:summary}, with the
properties of ClJ0152.7$-$1357 given for both the individual
subclusters and for the cluster as a whole, where appropriate. Cluster
properties derived using our alternative Einstein-de Sitter cosmology
are summarised in Table \ref{tab:summaryH50}.

\clearpage

\begin{deluxetable}{cccccccccccc}
\rotate
\tabletypesize{\scriptsize}
\tablewidth{22cm}
\tablecaption{\label{tab:summary}Summary of the measured and
inferred properties of the galaxy clusters ClJ1113.1$-$2615 and
ClJ0152.7$-$1357, based on the \Chandra observations discussed here
and assuming a \LCDM cosmology of $\Omega_{M}=0.3$
$(\Omega_\Lambda=0.7)$ and $H_0=70$\kmpspMpc. Where appropriate, the
properties of ClJ0152.7$-$1357 are listed for both the northern
(ClJ0152.7$-$1357N) and southern (ClJ0152.7$-$1357S) subclusters
individually, as well as for the cluster as a whole. The temperatures
quoted were derived from spectral fits with abundances frozen at $0.3$
solar.}
\tablehead{
\colhead{Cluster} & \colhead{Redshift} & \colhead{$T (keV)$} & \colhead{$L_{bol} (erg\s^{-1})$\tablenotemark{a}} & \colhead{$r_c ({\arcs})$} & \colhead{$\beta$} & \colhead{$e$} & \colhead{$r_v (Mpc)$} & \colhead{$M_g(r_v) (M_\odot)$} & \colhead{$M(r_v) (M_\odot)$ } & \colhead{$M_g/M (r_s)$} & \colhead{$M_g/M (r_v)$}}
\startdata
ClJ1113.1$-$2615 & 0.725 & $4.3^{+0.5}_{-0.4}$ & $3.3^{+0.3}_{-0.3}\times10^{44}$ & $14.6^{+1.2}_{-2.2}$ & $0.67^{+0.03}_{-0.05}$ & $0.2$ & $1.3\pm0.1$ & $3.0^{+0.4}_{-0.4}\times10^{13}$ & $4.3^{+0.8}_{-0.7}\times10^{14}$ & $0.05^{+0.01}_{-0.01}$ & $0.07^{+0.01}_{-0.01}$\\
ClJ0152.7$-$1357S & 0.833 & $5.2^{+1.1}_{-0.9}$ & $5.8^{+1.1}_{-0.9}\times10^{44}$ & $16.1^{+3.7}_{-2.6}$ & $0.66^{+0.08}_{-0.06}$ & $0.00$ & $1.4\pm0.1$ & $4.5^{+1.3}_{-1.1}\times10^{13}$ & $5.2^{+1.8}_{-1.4}\times10^{14}$ & $0.06^{+0.02}_{-0.01}$ & $0.09^{+0.04}_{-0.03}$ \\
ClJ0152.7$-$1357N & 0.833 & $5.5^{+0.9}_{-0.8}$ & $1.0\pm0.2\times10^{45}$ & $32.7^{+8.3}_{-4.3}$ & $0.73^{+0.13}_{-0.06}$ & $0.08$ & $1.4\pm0.1$ & $7.0^{+1.7}_{-1.5}\times10^{13}$ & $6.1^{+1.7}_{-1.5}\times10^{14}$ & $0.09^{+0.02}_{-0.01}$ & $0.12\pm0.04$ \\
ClJ0152.7$-$1357 & 0.833 & & $1.6\pm0.2\times10^{45}$ & & & & & & $1.1\pm0.2\times10^{15}$ \\
\enddata
\tablenotetext{a}{Bolometric luminosity computed from fluxes scaled out to the virial radius.}
\end{deluxetable}

\begin{deluxetable}{cccccccc}
\rotate
\tablewidth{19.5cm}
\tabletypesize{\scriptsize}
\tablecaption{\label{tab:summaryH50}
As for Table 2, but assuming an Einstein-de
Sitter cosmology of $H_0=50\kmpspMpc$ and $\Omega_M=1$
$(q_0=0.5)$.}
\tablehead{
\colhead{Cluster} & \colhead{$F(r_v)(0.5-2\keV)\flux$} & \colhead{$L_{bol} (erg\s^{-1})$\tablenotemark{a}} & \colhead{$r_v (Mpc)$} & \colhead{$M_g(r_v) (M_\odot)$} & \colhead{$M(r_v) (M_\odot)$} & \colhead{$M_g/M (r_s)$} & \colhead{$M_g/M (r_v)$}}
\startdata
ClJ1113.1$-$2615 & $5.7\pm0.5\times10^{-14}$ & $3.9^{+0.4}_{-0.3}\times10^{44}$ & $1.1\pm0.1$ & $2.8^{+0.4}_{-0.3}\times10^{13}$ & $3.6^{+0.7}_{-0.6}\times10^{14}$ & $0.06\pm0.01$ & $0.08\pm0.01$ \\
ClJ0152.7$-$1357S & $6.9^{+1.3}_{-1.0}\times10^{-14}$ & $6.7^{+1.2}_{-0.9}\times10^{44}$ & $1.1\pm0.1$ & $4.0^{+1.0}_{-0.8}\times10^{13}$ & $4.3^{+1.5}_{-1.2}\times10^{14}$ & $0.06^{+0.02}_{-0.01}$ & $0.09^{+0.04}_{-0.03}$ \\
ClJ0152.7$-$1357N & $1.1\pm0.2\times10^{-13}$ & $1.1\pm0.2\times10^{45}$ & $1.2\pm0.1$ & $6.3^{+1.3}_{-1.1}\times10^{13}$ & $5.0^{+1.4}_{-1.2}\times10^{14}$ & $0.10\pm0.02$ & $0.13\pm0.04$\\
ClJ0152.7$-$1357 & $1.8\pm0.2\times10^{-13}$ & $1.8\pm0.2\times10^{45}$ & & & $9.3^{+2.1}_{-1.7}\times10^{14}$ & &  \\
\enddata
\tablenotetext{a}{Bolometric luminosity computed from fluxes scaled out to the virial radius.}
\end{deluxetable}

\clearpage

We find ClJ1113.1$-$2615 to be a relaxed, hot and massive cluster at
$z=0.725$. The lack of pronounced substructure suggests that we are
observing this system some time after its formation, or any recent
mergers. On the other hand, the small but non-zero ellipticity of
the X-ray emission may be interpreted as a remnant of the last merger event.

ClJ0152.7$-$1357 is found to be a very massive, hot system at
$z=0.833$, which consists of two massive subclusters that are likely
in the process of merging. This scenario is supported by our dynamical
analysis of the system, which shows that the subclusters are likely to
be gravitationally bound, the X-ray contours, which show extensions
away from the hypothesised direction of the merger, and a displacement
between the viscous ICM and the essentially collisionless galaxies of
the southern subcluster, which suggests a motion of this subcluster in
a northerly direction toward the northern subcluster. It is difficult
to compare this observation with recent detailed simulations of
cluster mergers as they do not generally include galaxies.

Excess X-ray emission suggestive of a shock front is detected between
the subclusters. Numerical simulations of equal-mass mergers
\citep[\egc][]{roe96,rit02} show the formation of such a shock front
in the early stages of the merging process as the outer regions of the
subclusters meet and gas is driven out perpendicular to the merger
axis. The gas in the shocked region is further compressed and heated
as the merger continues, becoming detectable by its increased density
and temperature. The emission detected from the shock region in
ClJ0152.7$-$1357 is harder than the surrounding emission, which is
also consistent with the shock front scenario. 
The overall evidence is suggestive of a weak shock front in this
merging system, 
although the limited number of photons prevent a robust
detection. Future, deeper X-ray observations are needed to clarify the
situation.

\clearpage

\begin{figure}
\begin{center}
\plotone{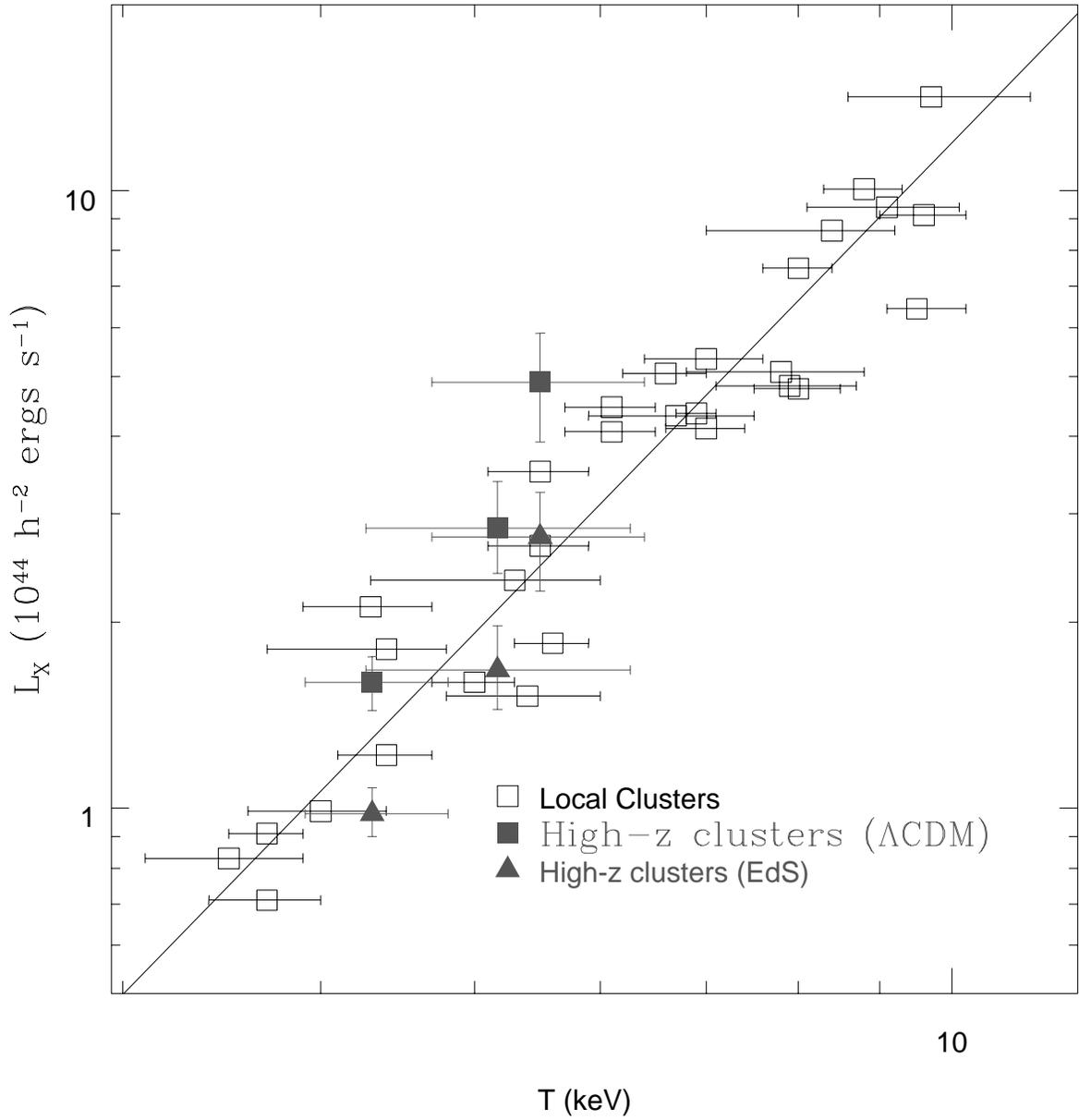}
\caption{\label{fig:LT}Bolometric X-ray luminosities and temperatures of the
clusters analysed here are plotted on the low-redshift L-T relation of
\citet{mar98a} in our two cosmologies. Our new data points are shown as filled squares (\LCDM)and filled triangles (Einstein-de Sitter), and
represent, from left to right, ClJ1113.1$-$2615, ClJ0152.7$-$1357S,
and ClJ0152.7$-$1357N.} 
\end{center}
\end{figure}

\clearpage

\subsection{Evolution of cluster scaling relations}
We now compare the X-ray properties of the three
clusters analysed here (we here treat the two subclusters of
ClJ0152.7$-$1357 separately) to those expected based on  low-z
scaling relations. We first examine the
luminosity-temperature (L-T) relation, using the temperatures
and luminosities of \citet{mar98a} corrected for cooling flow
contamination, which should be consistent with the quantities we
measure, as the high-redshift systems do not appear to host strong
cooling flows. The luminosities were derived in our two model
cosmologies, and are shown in Fig. \ref{fig:LT}. Previous work,
assuming an Einstein-de Sitter cosmology has tended to find little, or
no evolution in the L-T relation \citep[\egc][]{fai00}, and our new
data support this; the Einstein-de Sitter points are consistent with
the local L-T relation. This indicates that contamination by
unresolved point sources of observations made with earlier satellites
(e.g. by Fairley \etal) was not, on average, a very significant effect. 
Recent work, assuming a \LCDM cosmology
has found evidence for positive evolution of the L-T relation
\citep[\egc][]{arn02b,vik02}, i.e. clusters at high redshift are more
luminous for a given temperature (although \citet{hol02} find no significant evolution in this cosmology). Our data support this general trend of positive evolution,
albeit with low statistical significance, due to the few points
available.

\begin{figure}
\begin{center}
\plotone{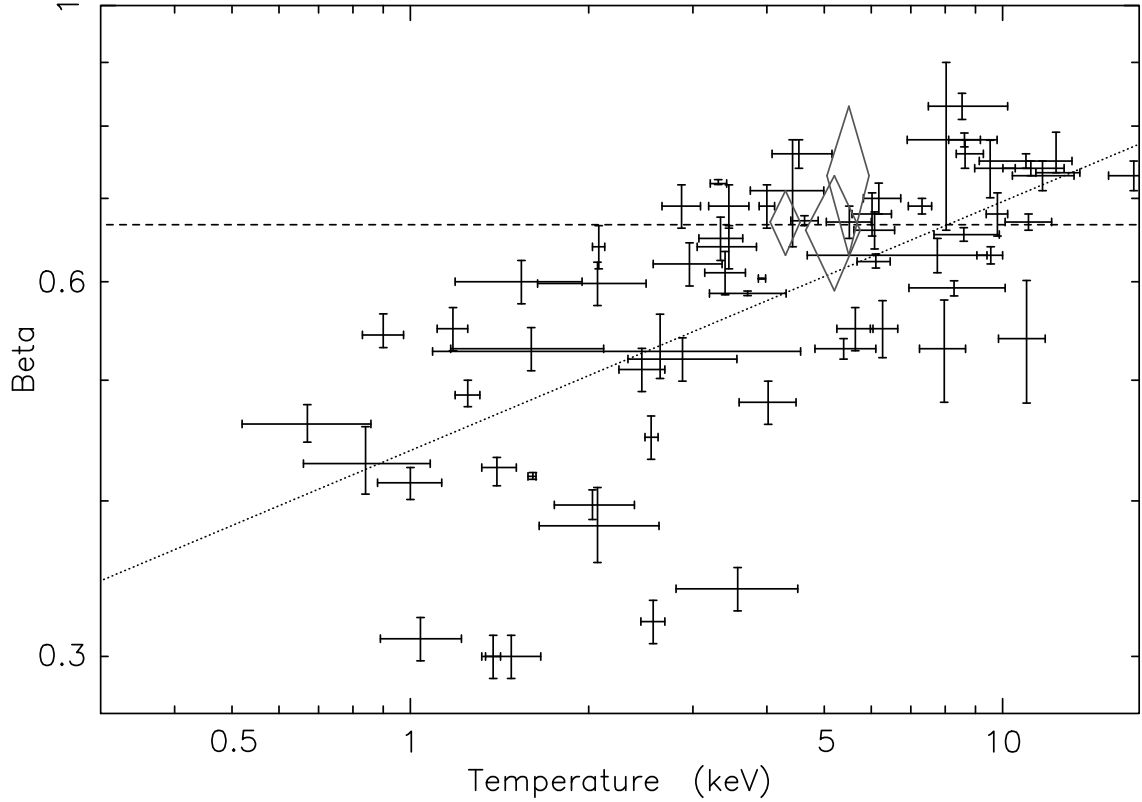}
\caption{\label{fig:betaT}$\beta$ values and temperatures of the
clusters analysed here plotted on the low-z $\beta-T$ relation of
\citet{san02}. Our new data are shown as diamonds, and represent, from
left to right, ClJ1113.1$-$2615, ClJ0152.7$-$1357S, and
ClJ0152.7$-$1357N. The solid line shows the best fit to the low-z
data, while the dashed line indicates the canonical value of
$\beta=2/3$.}
\end{center}
\end{figure}

The loci of our three clusters were also plotted on the low-redshift
$\beta$-T relation of \citet{san02} (Fig. \ref{fig:betaT}), which covers redshifts out to
$z\sim0.1$. Our three data points lie within the scatter around the
$\beta$-T relation, suggesting again that the clusters we observe at
redshifts $z\approx0.8$ have a similar distribution of gas to that
of systems in the local universe.

\clearpage

\clearpage

\begin{figure}
\begin{center}
\plotone{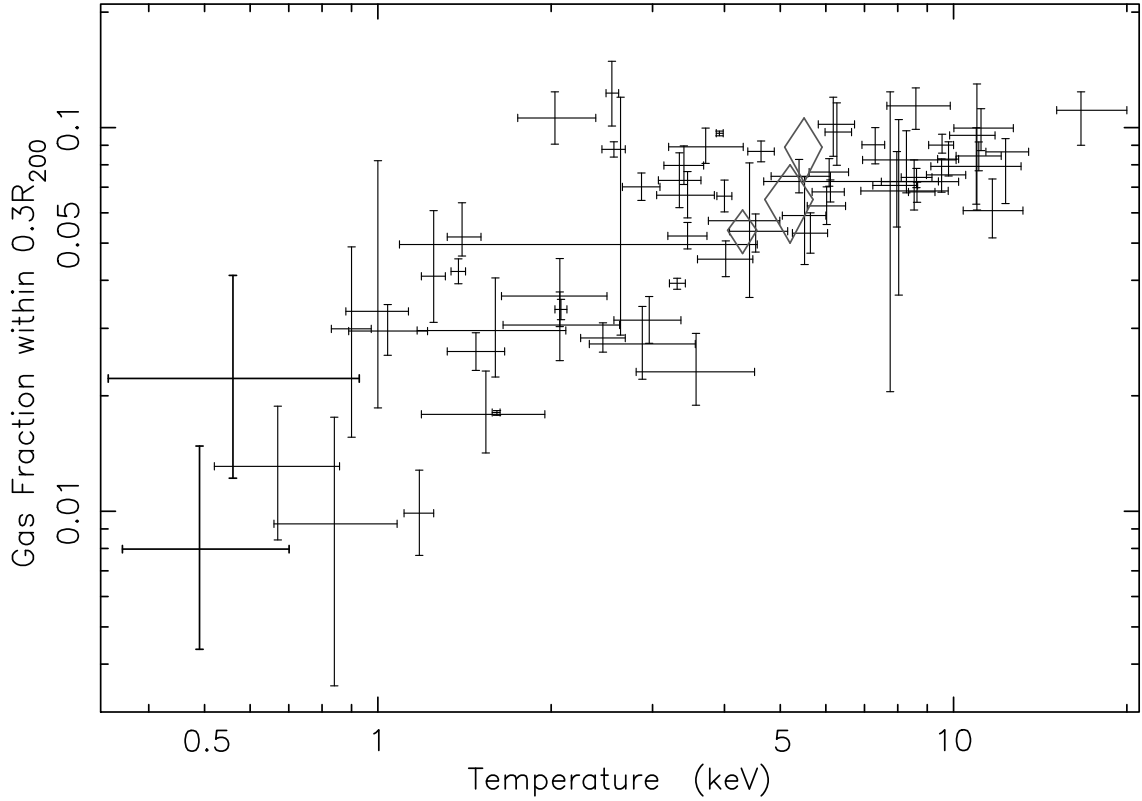}
\caption{\label{fig:gasfrac}The loci of the three clusters observed by
\Chandra discussed here (diamonds) plotted on the gas-mass fraction
versus temperature plot of \citet{san02} for low-redshift
clusters. For all clusters the gas-mass fraction was evaluated at
$0.3r_v$. The new data points represent, from left to right,
ClJ1113.1$-$2615, ClJ0152.7$-$1357S, and ClJ0152.7$-$1357N.}
\end{center}
\end{figure}

The gas mass fractions of all three clusters were measured within
their spectral radii, and total values were obtained by extrapolating
to their virial radii. Note that, since the X-ray emission was only
detected to $20-50\%$ of the virial radii, only around $20\%$ of the
estimated total gas mass is directly observed. To allow a comparison
of our results with with those from the low-redshift study of
\citet{san02} who compute cluster gas mass fractions at $0.3r_v$, we
also derived the gas mass fraction of our clusters at this radius
(slightly larger than our spectral radii) in our \LCDM
cosmology (this is consistent with the values of $H_0=70\kmpspMpc$
assumed by Sanderson and coworkers). The results are plotted along
with the data of \citet{san02} in Fig. \ref{fig:gasfrac}. The loci of
our high-redshift clusters do not deviate significantly from the low
redshift data, suggesting that, for this cosmology, the composition of
at least some galaxy clusters does not change significantly from $z\sim 0.8$ to $z\sim 0.1$. 

\clearpage

\clearpage

\subsection{The M-T relation}
Finally, we compare the mass-temperature relation found in the local
sample of \citet{san02} with the properties measured for our
high-redshift clusters, and for the intermediate redshift sample of
\citet{all01a}.  This intermediate redshift sample provides a useful
comparison, as it has secure mass measurements within $R_{2500}$
derived from \Chandra temperature profiles, and corroborated by weak-lensing measurements. The temperatures derived by those authors are unaffected
by the ACIS low energy QE degradation since the low energy absorption was
a free parameter in their analysis \citep{all02b}.

\begin{figure}
\begin{center}
\plotone{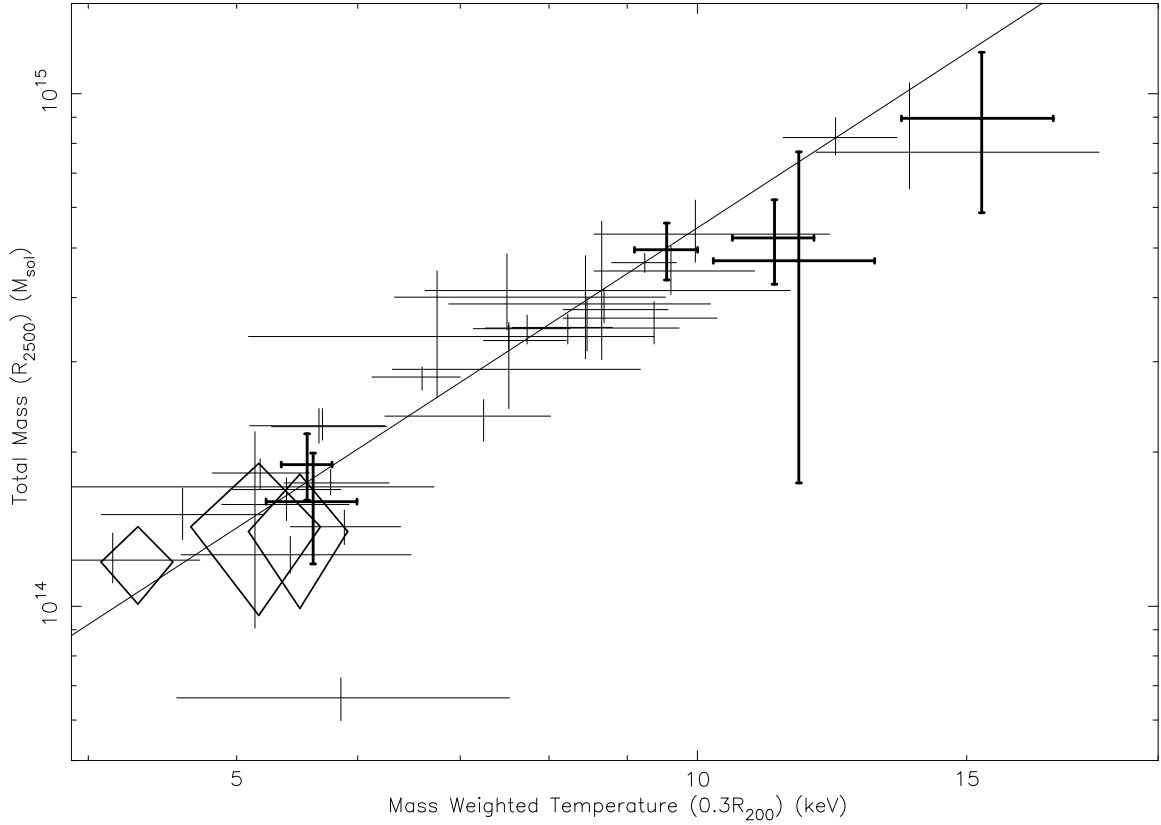}
\caption{\label{fig:MT}The loci of the three clusters analysed here are
plotted as diamonds on the low-z $M-T$ relation of \citet{san02}. 
The solid line is the best fit to the local data at T$>$4 keV and the
heavy, barred crosses are from the Chandra data of intermediate-redshift
clusters of \citet{all01a}.}
\end{center}
\end{figure}

Fig. \ref{fig:MT} shows the mass (within $R_{2500}$) of the hotter
($T>4\keV$) systems from the local sample plotted against their mass
weighted temperatures (measured within $0.3R_{200}$). The line shows
the best fitting power law to the local data points, of the form
$M=A_{2500}T^{\alpha}$, with
$A_{2500}=6.2^{+1.7}_{-1.4}\times10^{12}\Msol$ and
$\alpha=1.94\pm0.12$. The intermediate-redshift clusters (marked as
barred crosses) have their mass (within $R_{2500}$) plotted against
their mass weighted temperatures (also measured within
$R_{2500}$). Finally, our new high-redshift data are plotted as
diamonds, and we plot mass within $0.3R_{200}$ against
emission-weighted temperatures (under our assumption of isothermality,
the emission weighted temperature is the same as the mass weighted
temperature, and constant with radius, though the errors quoted do not
include any systematic uncertainties in the masses due to this
assumption). Note that $0.3R_{200}$ is very close to $R_{2500}$
\citep[\egc][]{san02}, so the different scales used are consistent.

The most important result is that little or no evolution of the M-T relation
is observed. In
order to investigate quantitatively the evolution in the M-T relation
we measured the mean offsets in mass of the high-redshift points, weighted by their
errors, from the local M-T relation. The mean offset factors, calculated for the
intermediate, and high-redshift samples are given in Table
\ref{tab:M_offset}. The errors quoted are the formal errors in the weighted
mean, based on the scatter of the data. For both samples the mean offset is consistent with zero,
a conclusion confirmed by reduced $\chi^2$ (and corresponding probability) values of $0.40$($0.75$) and $0.46$($0.84$)
for the intermediate and high-redshift samples respectively, when compared
to the local M-T relation.

The normalisation of the M-T relation is, however,  expected to be lower for
objects which formed at higher redshifts, because objects which form in
a denser universe will be hotter for a given mass ($T\propto M/R$). The evolution of
the normalisation can be modelled by a redshift dependent factor. A
cluster of a given mass forming at a redshift $z$ will have a
temperature given by $E(z)M\propto T^{3/2}$, where
$E(z)=(1+z)\sqrt{(1+z\Omega_M+\Omega_\Lambda/(1+z)^2-\Omega_\Lambda)}$.
 If we assume that the redshift of observation ($z_{obs}$) is the same as the redshift of
formation ($z_f$), we would expect 
the high-redshift clusters to lie below the local M-T relation by a mean factor 
$1/\overline{E(z)}$ also given in Table 
\ref{tab:M_offset} (for a \LCDM cosmology). In fact, increasing the observed masses 
by factors of $E(z)$ would place them above the local M-T relation, but
not very significantly so, given the reduced $\chi^2$ (and probability) values of $0.62$($0.71$) and $3.29$($0.02$) 
for the intermediate and high-redshift $E(z)M$ samples, when again compared
to the local M-T relation (given that the errors on the high-redshift sample are underestimates because of the assumption of isothermality).

If this lack of evolution in the normalisation of the M-T relation is confirmed
by further high-redshift cluster observations, then one interpretation is that
$z_{obs}\neq z_f$, but that clusters still reflect the conditions of the epoch at which
they formed. In this case the the normalisation of the local M-T relation
corresponds  to some mean redshift of formation of the local clusters. The
fact that the higher redshift clusters lie on the same, or a similar, M-T relation
suggests that their
redshift of formation is similar to that of the local systems.

An alternative, and perhaps more realistic, interpretation would be that 
clusters continuously grow in mass, rather than forming in one major
merger event identified with a formation epoch. In this case the properties of clusters more closely correspond
to the epoch at which they are observed, and less evolution in the M-T relation
is predicted \citep{voi00,mat01}. A more detailed comparison awaits better data.

\clearpage

\begin{deluxetable}{ccc}
\tablecaption{\label{tab:M_offset}Mean weighted fractional offset factors in mass 
($\overline{dM}$) from
the local M-T relation of the \citet{san02} data above $4\keV$,
and the predicted offset, $1/\overline{E(z)}$, based on the assumption
of a single redshift of formation for each cluster that is the same as
the observation redshift.
The local offset excludes AWM7, the point far below the relation at
$\approx5.8\keV$.}
\tablewidth{0pt}
\tablehead{
\colhead{Redshift Range} & \colhead{$\overline{dM}$} & \colhead{$1/\overline{E(z)}$}}
\startdata
$<0.1$ & $0.99\pm0.03$ &   \\
$0.1-0.45$ & $0.95\pm0.05$ & $0.85$ \\
$0.72-0.83$ & $1.03\pm0.08$ & $0.64$ \\
\enddata
\end{deluxetable}

\clearpage

\clearpage

\section{Conclusions}
We have analysed \Chandra X-ray observations of two massive
high-redshift clusters of galaxies: CLJ1113.1$-$2615 (z=0.725) and
ClJ0152.7$-$1357 (z=0.833). ClJ1113.1$-$2615 appears largely relaxed,
whereas ClJ0152.7$-$1357 is likely to be in the early phase of an
equal-mass merger of sufficient mass to create the Coma cluster; 
at 80\% confidence the system is gravitationally bound.

Treating ClJ0152.7$-$1357 as two independent subclusters,
we find the hot gas in the clusters studied to have very similar
global properties to that in massive clusters at low redshift.
The metallicity, gas fraction and gas density profile slope are all 
consistent with the values found in local clusters of similar temperature,
suggesting that the gas was in place, and containing its metals,
at $z=0.8$. These properties point towards an early
epoch of assembly of at least some massive clusters.

We measure the M-T relation at $z=0.8$ for the first time,
and find  no significant evolution of the normalisation.
This behaviour  is more consistent with models in which clusters are 
continually evolving, rather than those in which each cluster reflects
the conditions at a single formation epoch. It also supports the
use of the X-ray temperature function at high redshifts to constrain 
cosmological parameters, especially $\Omega_m$.

Finally, our data support the general consensus of little, or no evolution in
the L-T relation under the assumption of an Einstein-de Sitter
cosmology, and are consistent with recent suggestions of positive
evolution in a \LCDM cosmology.


\section{Acknowledgements}
We thank M. Arnaud and P. Thomas for useful discussions.
LRJ acknowledges the UK PPARC for support during part of this
work. BJM is supported by a PPARC postgraduate studentship. HE and EP
gratefully acknowledge financial support from SAO grant GO0-1071A.

\appendix
\section{Effect of ACIS QE degradation on spectra}\label{app1}
In this appendix we examine the effect of the ACIS QE degradation due
to hydrocarbon contamination on the spectra analysed here, and the
methods of correcting for this. There are currently three similar
methods for correcting for the problem: The ACISABS model for XSPEC
(http://\-www.astro.psu.edu/\-users/\-chartas/\-xcontdir/), the same model, implemented for Sherpa
(http://\-cxc.harvard.edu/\-ciao/\-threads/\-sherpa\_acisabs/\-), and corrarf,
a Fortran code that corrects the ARF using the ACISABS absorption
model (http://\-cxc.harvard.edu/\-cal/\-Links/\-Acis/\-acis/\-Cal\_prods/\-qeDeg/\-corrarf.tar.gz). All three methods use the same
contamination rate, which has been derived from calibration
observations, and assume the same composition of the contaminant.

We take the spectrum of ClJ1113.1$-$2615 as a test case, as this is
the best quality spectrum of the three discussed here. The spectrum
was fit with an absorbed \MEKAL model, uncorrected, and with each of
the three correction methods. As the contamination increases with
time, the correction methods require the number of days between
\Chandra launch and the observation, to normalise the absorption. The
observation of ClJ1113.1$-$2615 was taken $387$ days after launch. The
fits were performed in two energy ranges, $0.5-8\keV$ and $1-8\keV$,
and the column density was either frozen at the Galactic value, or
allowed to fit. The best fitting temperatures and column densities are
given in Table \ref{tab:acisabs}. The spectral parameters found for
the three correction methods were completely consistent, so we just
give those for XSPEC ACISABS for clarity. The \MEKAL abundance was
frozen at $0.3\Zsol$ for simplicity.

\clearpage

\begin{deluxetable}{ccccccc}
\rotate
\tabletypesize{\small}
\tablecaption{\label{tab:acisabs}Best fitting spectral parameters with and without correcting for the \Chandra low energy QE degradation. Hydrogen columns marked with (f) were frozen at the galactic value during the fit.}
\tablewidth{0pt}
\tablehead{
\colhead{Correction} & \colhead{Range ($\keV$)} & \colhead{nH ($10^{22}\pcmsq$)} & \colhead{kT ($\keV$)} & \colhead{Range ($\keV$)} & \colhead{nH ($10^{22}\pcmsq$)} & \colhead{kT ($\keV$)}}
\startdata
None & $0.5-8\keV$ & $0.054$(f) & $5.2^{+0.7}_{-0.6}$ & $1-8\keV$ & $0.054$(f) & $4.7^{+0.6}_{-0.6}$ \\
None & $0.5-8\keV$ & $0.12^{+0.04}_{-0.04}$ & $4.2^{+0.7}_{-0.6}$ & $1-8\keV$ & $0.062^{+0.142}_{-0.062}$ & $4.7^{+0.7}_{-1.1}$ \\
ACISABS & $0.5-8\keV$ & $0.054$(f) & $4.4^{+0.7}_{-0.5}$ & $1-8\keV$ & $0.054$(f) & $4.4^{+0.6}_{-0.5}$ \\
ACISABS & $0.5-8\keV$ & $0.071^{+0.04}_{-0.04}$ & $4.3^{+0.7}_{-0.7}$ & $1-8\keV$ & $0.054^{+0.109}_{-0.054}$ & $4.4^{+0.8}_{-0.8}$ \\
\enddata
\end{deluxetable}

\clearpage

As the QE degradation is most severe below $1\keV$ we would expect the
spectra fitted above this energy to give similar results, which is
indeed the case. When the fit is extended below $1\keV$, however, we
find the the temperature is overestimated in the uncorrected spectra
by $\approx15\%$ when the absorbing column is fixed at the Galactic
value. On the other hand, if the absorbing column is allowed to fit,
the temperature found is generally accurate, while the absorption is
significantly overestimated. It is reassuring, though, that all of the
corrected spectral fits give consistent temperatures in both energy
ranges, and agree well with the uncorrected spectra above $1\keV$. The
best fitting column densities given by the corrected spectral fits is
also consistent with the Galactic value.

\bibliographystyle{apj}
\bibliography{clusters}

\begin{thebibliography}{43}
\expandafter\ifx\csname natexlab\endcsname\relax\def\natexlab#1{#1}\fi

\bibitem[{Allen {et~al.}(2002)Allen, , Fabian, Schmidt, \& H.}]{all02b}
Allen, S.~W., , Fabian, A.~C., Schmidt, R.~W., \& H., E. 2002, astro-ph/0208394

\bibitem[{Allen {et~al.}(2001)Allen, Schmidt, \& Fabian}]{all01a}
Allen, S.~W., Schmidt, R.~W., \& Fabian, A.~C. 2001, MNRAS, 328, L37

\bibitem[{{Allen} {et~al.}(2002){Allen}, {Schmidt}, \& {Fabian}}]{all02a}
{Allen}, S.~W., {Schmidt}, R.~W., \& {Fabian}, A.~C. 2002, MNRAS, 334, L11

\bibitem[{Arnaud(1996)}]{arn96}
Arnaud, K.~A. 1996, ASP Conf. Series, 101, 17

\bibitem[{Arnaud {et~al.}(2002{\natexlab{a}})Arnaud, Aghanim, \&
  Neumann}]{arn02a}
Arnaud, M., Aghanim, N., \& Neumann, D.~M. 2002{\natexlab{a}}, A\&A, 389, 1A

\bibitem[{Arnaud {et~al.}(2002{\natexlab{b}})Arnaud, Aghanim, \&
  Neumann}]{arn02b}
---. 2002{\natexlab{b}}, A\&A, 390, 27A

\bibitem[{{Birkinshaw} {et~al.}(1991){Birkinshaw}, {Hughes}, \&
  {Arnaud}}]{bir91}
{Birkinshaw}, M., {Hughes}, J.~P., \& {Arnaud}, K.~A. 1991, ApJ, 379, 466

\bibitem[{{Borgani} {et~al.}(2001){Borgani}, {Rosati}, {Tozzi}, {Stanford},
  {Eisenhardt}, {Lidman}, {Holden}, {Della Ceca}, {Norman}, \&
  {Squires}}]{bor01}
{Borgani}, S., {Rosati}, P., {Tozzi}, P., {Stanford}, S.~A., {Eisenhardt},
  P.~R., {Lidman}, C., {Holden}, B., {Della Ceca}, R., {Norman}, C., \&
  {Squires}, G. 2001, \apj, 561, 13

\bibitem[{Cash(1979)}]{cas79}
Cash, W. 1979, ApJ, 228, 939

\bibitem[{Cavaliere \& Fusco-Femiano(1976)}]{cav76}
Cavaliere, A. \& Fusco-Femiano, R. 1976, A\&A, 49, L137

\bibitem[{Della~Ceca {et~al.}(2000)Della~Ceca, Scaramella, Gioia, Rosati, \&
  Squires}]{del00}
Della~Ceca, R., Scaramella, R., Gioia, I.~M., Rosati, F., \& Squires, G. 2000,
  A\&A, 353, 498

\bibitem[{Demarco {et~al.}(2002)Demarco, Rosati, {et~al.}}]{dem02}
Demarco, R., Rosati, P., {et~al.} 2002, A\&A, in preparation

\bibitem[{Dickey \& Lockman(1990)}]{dic90}
Dickey, J.~M. \& Lockman, F.~J. 1990, ARA\&A, 28, 215

\bibitem[{Ebeling(2002)}]{ebe02}
Ebeling, H. 2002, MNRAS, submitted

\bibitem[{Ebeling {et~al.}(2000)Ebeling, Jones, Perlman, Scharf, Horner,
  Wegner, Malkan, Fairley, \& Mullis}]{ebe00}
Ebeling, H., Jones, L.~R., Perlman, E., Scharf, C., Horner, D., Wegner, G.,
  Malkan, M., Fairley, B., \& Mullis, C.~R. 2000, ApJ, 534, 133

\bibitem[{Ettori {et~al.}(2003)Ettori, Tozzi, \& Rosati}]{ett03}
Ettori, S., Tozzi, P., \& Rosati, P. 2003, A\&A, in press, (astroph/0211335)

\bibitem[{Evrard {et~al.}(1996)Evrard, Metzler, \& Navarro}]{evr96}
Evrard, A.~E., Metzler, C.~A., \& Navarro, J.~F. 1996, ApJ, 469, 494

\bibitem[{Fabian(1994)}]{fab94b}
Fabian, A.~C. 1994, ARA\&A, 32, 277

\bibitem[{Fairley {et~al.}(2000)Fairley, Jones, Scharf, Ebeling, Perlman,
  Horner, Wegner, \& Malkan}]{fai00}
Fairley, B.~W., Jones, L.~R., Scharf, C., Ebeling, H., Perlman, E., Horner, D.,
  Wegner, G., \& Malkan, M. 2000, MNRAS, 315, 669

\bibitem[{Henry(2000)}]{hen00}
Henry, J.~P. 2000, ApJ, 533

\bibitem[{Henry \& Arnaud(1991)}]{hen91}
Henry, J.~P. \& Arnaud, K.~A. 1991, ApJ, 372, 410

\bibitem[{{Holden} {et~al.}(2002){Holden}, {Stanford}, {Squires}, {Rosati},
  {Tozzi}, {Eisenhardt}, \& {Spinrad}}]{hol02}
{Holden}, B.~P., {Stanford}, S.~A., {Squires}, G.~K., {Rosati}, P., {Tozzi},
  P., {Eisenhardt}, P., \& {Spinrad}, H. 2002, ApJ, 124, 33

\bibitem[{Hughes {et~al.}(1996)Hughes, Birkinshaw, \& Huchra}]{hug95}
Hughes, J.~P., Birkinshaw, M., \& Huchra, J.~P. 1996, ApJ, 448, L93

\bibitem[{Jones {et~al.}(1998)Jones, Scharf, Ebeling, Perlman, Wegner, Malkan,
  \& Horner}]{jon98a}
Jones, L.~R., Scharf, C., Ebeling, H., Perlman, E., Wegner, G., Malkan, M., \&
  Horner, D. 1998, ApJ, 495, 100

\bibitem[{Joy {et~al.}(2001)Joy, LaRoque, Grego, Carlstrom, Dawson, Ebeling,
  Holzapfel, Nagai, \& Reese}]{joy01}
Joy, M., LaRoque, S., Grego, L., Carlstrom, J.~E., Dawson, K., Ebeling, H.,
  Holzapfel, W.~L., Nagai, D., \& Reese, E.~D. 2001, ApJ, 551, L1

\bibitem[{Kaastra \& Mewe(1993)}]{kaa93}
Kaastra, J.~S. \& Mewe, R. 1993, aasupp, 97, 443

\bibitem[{Liedahl {et~al.}(1995)Liedahl, Osterheld, \& Goldstein}]{lie95}
Liedahl, D.~A., Osterheld, A.~L., \& Goldstein, W.~H. 1995, ApJ, 438, L115

\bibitem[{Markevitch(1998)}]{mar98a}
Markevitch, M. 1998, ApJ, 504, 27

\bibitem[{Markevitch(2001)}]{mar00b}
---. 2001, ACIS background (http://hea-www.harvard.edu/~maxim/axaf/acisbg/)

\bibitem[{Markevitch {et~al.}(2002)Markevitch, Gonzalez, Vikhlinin, Murray,
  Forman, Jones, \& Tucker}]{mar02}
Markevitch, M., Gonzalez, L., Vikhlinin, A., Murray, S., Forman, W., Jones, C.,
  \& Tucker, W. 2002, ApJ, 567, L27

\bibitem[{Markevitch {et~al.}(2000)Markevitch, Ponman, Nulsen, Bautz, Burke,
  David, Davis, Donnelly, Forman, Jones, Kaastra, Kellogg, Kim, Kolodziejczak,
  Mazzotta, Pagliaro, Patel, Van~Speybroeck, Vikhlinin, Vrtilek, Wise, \&
  Zhao}]{mar00c}
Markevitch, M., Ponman, T.~J., Nulsen, P. E.~J., Bautz, M.~W., Burke, D.~J.,
  David, L.~P., Davis, D., Donnelly, R.~H., Forman, W.~R., Jones, C., Kaastra,
  J., Kellogg, E., Kim, D.-W., Kolodziejczak, J., Mazzotta, P., Pagliaro, A.,
  Patel, S., Van~Speybroeck, L., Vikhlinin, A., Vrtilek, J., Wise, M., \& Zhao,
  P. 2000, ApJ, 541, 542

\bibitem[{{Mathiesen}(2001)}]{mat01}
{Mathiesen}, B.~F. 2001, MNRAS, 326, L1

\bibitem[{McNamara {et~al.}(2001)McNamara, Wise, Nulsen, David, Carilli,
  Sarazin, O'Dea, Houck, Donahue, Baum, Voit, O'Connell, \& Koekemoer}]{mcn01}
McNamara, B.~R., Wise, M.~W., Nulsen, P. E.~J., David, L.~P., Carilli, C.~L.,
  Sarazin, C.~L., O'Dea, C.~P., Houck, J., Donahue, M., Baum, S., Voit, M.,
  O'Connell, R.~W., \& Koekemoer, A. 2001, ApJ, 562, L149

\bibitem[{Perlman {et~al.}(2002)Perlman, Horner, Jones, Scharf, Ebeling,
  Wegner, \& Malkan}]{per02}
Perlman, E., Horner, D., Jones, L.~R., Scharf, C., Ebeling, H., Wegner, G., \&
  Malkan, M. 2002, ApJS, in press

\bibitem[{Ritchie \& Thomas(2002)}]{rit02}
Ritchie, B.~W. \& Thomas, P.~A. 2002, MNRAS, 329, 675

\bibitem[{Roettiger {et~al.}(1996)Roettiger, Burns, \& Loken}]{roe96}
Roettiger, K., Burns, J.~O., \& Loken, C. 1996, ApJ, 473, 651

\bibitem[{Romer {et~al.}(2000)Romer, Nichol, Holden, Ulmer, Pildis, Merrelli,
  Adami, Burke, Collins, Metevier, Kron, \& Commons}]{rom00}
Romer, A.~K., Nichol, R.~C., Holden, B.~P., Ulmer, M.~P., Pildis, R.~A.,
  Merrelli, A.~J., Adami, C., Burke, D.~J., Collins, C.~A., Metevier, A.~J.,
  Kron, R.~G., \& Commons, K. 2000, ApJS, 126, 209

\bibitem[{Rosati {et~al.}(1998)Rosati, Della~Ceca, Norman, \& Giaconni}]{ros98}
Rosati, P., Della~Ceca, R., Norman, C., \& Giaconni, R. 1998, ApJ, 492, L21

\bibitem[{Sanderson {et~al.}(2002)Sanderson, Ponman, Finoguenov, Lloyd-Davies,
  \& Markevitch}]{san02}
Sanderson, A. J.~R., Ponman, T.~J., Finoguenov, A., Lloyd-Davies, E.~L., \&
  Markevitch, M. 2002, MNRAS, submitted

\bibitem[{Scharf {et~al.}(1997)Scharf, Jones, Ebeling, Perlman, Malkan, \&
  Wegner}]{sch97}
Scharf, C., Jones, L.~R., Ebeling, H., Perlman, E., Malkan, M., \& Wegner, G.
  1997, ApJ, 477, 79

\bibitem[{Schindler {et~al.}(2001)Schindler, Castillo-Morales, De~Filippis,
  Schwope, \& Wambsganss}]{sch01a}
Schindler, S., Castillo-Morales, A., De~Filippis, E., Schwope, A., \&
  Wambsganss, J. 2001, aanda, 376, L27

\bibitem[{Vikhlinin {et~al.}(2002)Vikhlinin, VanSpeybroeck, Markevitch, Forman,
  \& Grego}]{vik02}
Vikhlinin, A., VanSpeybroeck, L., Markevitch, M., Forman, W.~R., \& Grego, L.
  2002, astro-ph/0207445

\bibitem[{{Voit}(2000)}]{voi00}
{Voit}, G.~M. 2000, ApJ, 543, 113

\end{thebibliography}

\end{document}